# ESCAPE

Preparing Forecasting Systems for the Next generation of Supercomputers

# D3.5 Projections of achievable performance for Weather & Climate Dwarfs, and for entire NWP applications, on hybrid architectures

Dissemination Level: Public

This project has received funding from the European Union's Horizon 2020 research and innovation programme under grant agreement No 67162

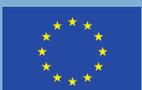 Funded by the European Union

Co-ordinated by

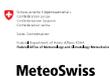 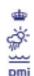 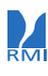 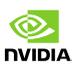 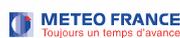 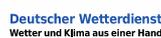 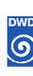 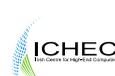 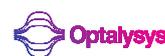 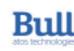 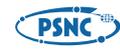 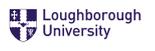

# ESCAPE

**Energy-efficient Scalable Algorithms for Weather Prediction at Exascale**

Author Michał Kulczewski, Marek Błażewicz, Sebastian Ciesielski

Date 10/12/2018

Table of Contents

# Figures

# Tables

# 1 Executive Summary

This deliverable contains the description of the performance and energy models for the selected Weather & Climate dwarfs for different hardware architectures, multinode with GPU accelerators in particular. Presented performance models are extension to model provided in Deliverable 3.2. With some further enhancements, they are incorporated in the DCworms simulator (see Deliverable 4.1 for details). In particular, extended models allow to predict computational and energy performance on different architectures: single and multinodes, equipped with CPUs and GPUs accelerators. This allows to provide feasible performance projection at system scale.

# 2 Introduction

## 2.1 Background

ESCAPE stands for Energy-efficient Scalable Algorithms for Weather Prediction at Exascale. The project develops world-class, extreme-scale computing capabilities for European operational numerical weather prediction and future climate models. ESCAPE addresses the ETHP4HPC Strategic Research Agenda "Energy and resiliency" priority topic, promoting a holistic understanding of the energy-efficiency for extreme-scale applications using heterogeneous architectures, accelerators and special compute units by:
- Defining and encapsulating the fundamental algorithmic building blocks underlying weather and climate computing;
- Combining cutting-edge research and algorithm development for use in extreme-scale, high-performance computing applications, minimizing time- and cost-to-solution;
- Synthesizing the complementary skills of leading weather forecasting consortia, university research, high-performance computing centres, and innovative hardware companies.

ESCAPE is funded by European Commissions' Horizon 2020 funding framework under the Future and Emerging Technologies – High-Performance Computing call for research and innovation action issued in 2014.

## 2.2 Scope of this deliverable

### 2.2.1 Objectives of this deliverable

The aim of this deliverable is to present how the internal model workflow can be optimally realised for the whole application on hybrid architectures. A bunch of performance tests for different Weather & Climate Dwarfs and NWP application on different architectures are to be provided, using the DCworms simulator and performance models. It will allow for finding the best model workflow and architecture for given application, with respect to the proposed performance and energy efficiency metrics, e.g. time-to-solution, energy-to-solution, trade-off between energy consumption and performance.

### 2.2.2 Work performed in this deliverable

This deliverable provides extension to the performance model proposed in Deliverable 3.2. The single CPU model has been extended to support multinode environment, where Message Passing Interface (MPI) is used to exchange information between parallel parts of the application running on different nodes. This model has been applied to Spherical Harmonics and BiFFT dwarfs. Both are used in the operational weather prediction, in global and Limited Area Models respectively.

Performance model for accelerators has been provided. We focused on three GPU architectures: Fermi, Kepler and Maxwell. The model has been applied to ACRANEB2 radiation dwarf.

To project energy efficiency, adequate model for single- and multinode machines equipped with CPU and GPU is proposed and tested for Spherical Harmonics, BiFFT and ACRANEB2 dwarfs.

The last step was to provide means to project performance at system scale. To this end, we combine aforementioned models for performance and energy efficiency into the simulator. DCworms allows for modelling the whole data centre load with different applications running. In order to stay with project requirements, we provided further enhancements to support modelling NWP application as a workflow of different dwarfs that works together, running on homogeneous or heterogeneous platforms, e.g. CPUs and GPUs. As an example, a NWP workflow of BiFFT and ACRANEB2 dwarfs is presented, running on system equipped with CPU (BiFFT) and GPU (ACRANEB2).

### 2.2.3 Deviations and counter measures

The performance tests and models concerned initially different type of accelerators. During the course of the project, Intel announced to remove next generation of Xeon Phi product from its roadmap, thus this product line for HPC market is discontinued. Therefore, we focus on GPU accelerators only. However, performance and energy models, as well as DCworms simulator, are capable of handling hardware accelerators such as Xeon Phi in the future or any other hardware architecture yet to come.

## 3 Roofline model for GPU

### 3.1 State-of-the art

Nvidia GPU architecture is very sophisticated and complex massively parallel environment called by creators: SIMT (Single Instruction Multiple Thread). A simplified version of this architecture is presented in Figure 1.

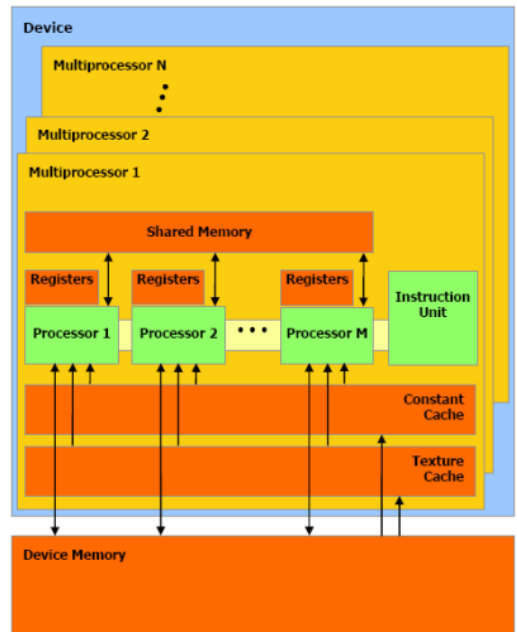

*Figure 1 Nvidia GPU architecture*

The GPU consists of many multiprocessors (abbreviated to SM), and each of them contains multiple processing units. Each multiprocessor is able to perform multiple instructions per clock cycle - the number differs depending on the generation and model of the device. Performance of SM may vary also due to the type of operation and the type of processed datum. The biggest difference is noted for double precision, for which most of the consumer devices have very poor peak performance. Only devices designed for HPC have better support for double precision operations, but the cost of such devices is significantly higher.
Computations deployed on a GPU device are called kernels. By definition, kernels are small pieces of parallelised computations that need to be divided into logical blocks of threads. One block cannot be run on multiple multiprocessors, while one multiprocessor can handle multiple blocks at the same time. Multiprocessor divides single block into warps – groups of 32 threads, see Figure 2.

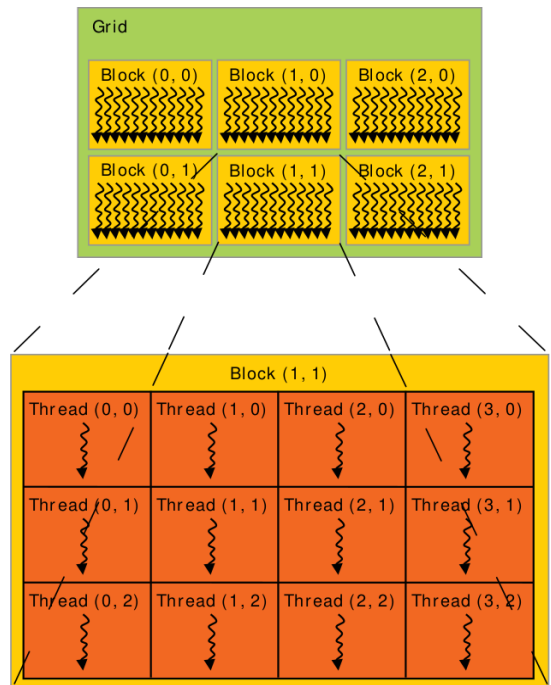

*Figure 2 Architecture of deployed kernel*

It is very important to utilise the full potential of the massively parallel architecture of GPUs. The most important optimisations are:

- All threads within a warp have to perform the same instruction since there is only one instruction unit per warp. If the threads diverge, the instruction scheduler needs to repeat instructions until all paths are finished. According to Nvidia documentation the divergence is indicated by control flow instructions like loops and conditionals.
- Memory should be accessed in an organised way. Global memory, the main storage for computations on GPU, is particularly prone to data access pattern. Proper usage has the main impact on the kernel performance. The threads of a warp should coalesce memory transactions into as few as one when certain access conditions are met in order to utilise the full potential of the GPU. However, the optimal memory access patterns differ between GPU generations. Not fulfilling them might result in fetched data being multiplication of the data actually needed in the computations. Shared memory – co-utilised between threads within a block – needs to be accessed in a proper way to avoid bank conflicts. The number of the banks as well as strategies to avoid conflicts differ with the change of the generation. These properties need to be taken into consideration by a developer while writing the code and deploying it on different GPU architectures.
- All GPU computational resources need to be fed with computations to utilise their full potential. First of all, the size of computational problem should have enough parallel and independent operations, so that for each computational unit there are at least few threads. Secondly, kernels cannot utilise too many local resources per block – like shared memory or registers. Not fulfilling this scheme results in processing units being idle, waiting for data to be fetched instead of overlapping data transfer with computations.

- Extensive synchronisation between threads within a block (e.g. calling function __syncthreads) or within the whole kernel (e.g. atomic operations) can drastically reduce the overall performance.

Understanding these mechanisms within GPU device is crucial to the optimal performance. When comparing to typical x86 processing units, GPUs have much better performance in transferring data from and to chip, whereas ratio of peak bandwidth to peak computational performance is typically higher on CPUs. This means that full potential can be exploited only for computationally intensive portions of code. Requirement for large number of independent and structured computations is very important factor for the overall success. Last but not least, transferring data between CPU and GPU memory, which is an order of magnitude slower than fetching data from global memory, is the performance bottleneck. Large data movements to and from GPU between executions of kernels can ruin benefits of using this specialised accelerator.

### 3.2 ACRANEB2

The radiation schemes in numerical weather prediction and climate models take up a considerable amount of the overall running time. Thus, a better utilisation of the radiation schemes on future extremely parallelised multithreaded CPUs and GPUs is in demand. ACRANEB2 dwarf is a radiation scheme made for short range weather models. This dwarf calculates atmospheric heating rates and specific downward surface fluxes for both shortwave and longwave radiation. There are different versions available, for Intel Xeon Phi and GPU architectures in particular. The proposed GPU model is based on ACRANEB2 analysis. This GPU dwarf is a standalone version of the *transt3* subroutine, described in details in Deliverable D1.2. This subroutine takes more than 80% of the total running time of ACRANEB2. Details on implementation and input variables to this dwarf can be found in Deliverable D1.2. Details on GPU optimisation are discussed in [J. W. Poulsen and P. Berg].

### 3.3 Code profiling

The code is written in Fortran 90 and ported to CUDA with the use of OpenACC. OpenACC is a standard developed by Cray, CAPS, Nvidia and PGI. The primary mode of programming in OpenACC is by using directives (pragmas). The developer tags code with proper pragmas informing compiler about the most intense fragments of the code that can be executed in parallel. The compiler and runtime library further decide how to port code onto GPU and how to structure computational kernels (division of the computational grid onto blocks). OpenACC also decides which mechanisms to use within CUDA application. To this end, we have used PGI compiler (version 17.4) which utilises CUDA 8.0.

The ACRANEB2 dwarf consists of seven separate kernels. Each kernel has different characteristics. The kernels were profiled separately with the use of Nvidia CUDA profiler. The benchmarks were performed on 3 different GPUs: Teslam2070-q, Tesla K20m, GeForce 970. Each of the devices is representing different microarchitecture (generation): Fermi, Kepler and Maxwell respectively. The first two are dedicated for HPC, while the last device is addressed to gaming industry. Devices are sorted from the oldest microarchitecture to the newest one. Profiling was performed for a variety of sizes of the computational grid, from 5x5 up to 200x200. The number of vertical levels was fixed to 80.

The Nvidia CUDA profiler analyses kernels in respect to fulfilling the CUDA-specific optimisation schemes, recording multiple counters. The most important are information about the number of performed instructions of each type, the level of achieved parallelism and efficiency of the memory access. In a simplified version, it presents achieved computational performance and bandwidth. As an outcome (report) of the profiling user also receives more general information deduced from the counters. The most informative is the chart that presents how much time each of the types of instructions took. The exemplary chart is presented in Figure 3, which depicts that GPU was inactive for the most of the time. According to information provided in the profiler report, inactivity is caused by the fact that thread was predicated or inactive due to divergence. Unfortunately, the inactivity was reported as the main issue, taking the most of the execution time for all the kernels on all tested GPUs. We believe is it because OpenACC is not capable to fully understand the underlying code and thus to fully optimise it for massively parallel GPU architectures. Because the profiler has not reported any divergence within the kernels, we believe that inactivity is caused mostly by extensive synchronisation and poor overlapping between computations and data transfer. Due to large impact on the kernel performance, the inactivity time was taken into consideration while modelling the total time of execution. Further investigation of hand-optimised kernels is required to enhance proposed model.

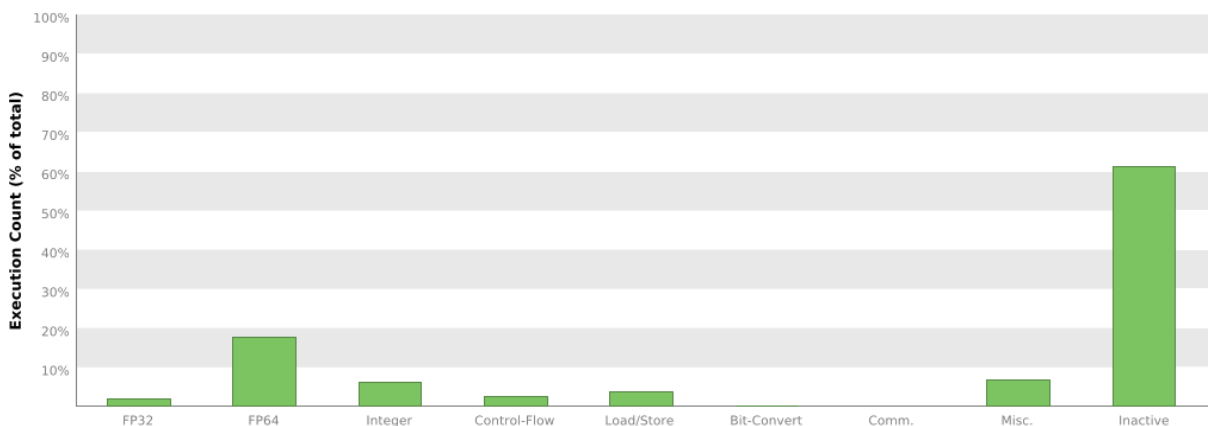

Figure 3 Exemplary chart representing the time spent on different type of instructions in kernel acraneb_transt3_649_gpu on GeForce 970 GPU

From the analysis given by Nvidia CUDA profiler we observed that the majority of computations is performed using double precision. That is why we treat these operations as *W* (amount of work). We treat requested global memory load and store as *Q*. These memory accesses describe what amount of data was actually needed by the kernel. The rooflines have been computed with respect to these assumptions.

Rooflines based on real benchmarks for 200x200x80 computational domain are presented below. Figures 4 - 6 present charts for individual kernels and whole dwarf for different GPUs. Individual kernels are presented only for one size of the domain, since it does not have major impact on performance. Figures 7 - 9 present roofline for dwarfs running over all benchmarked sizes of domains. The same results are also presented in tabular form in Tables 1 - 6 for convenience.

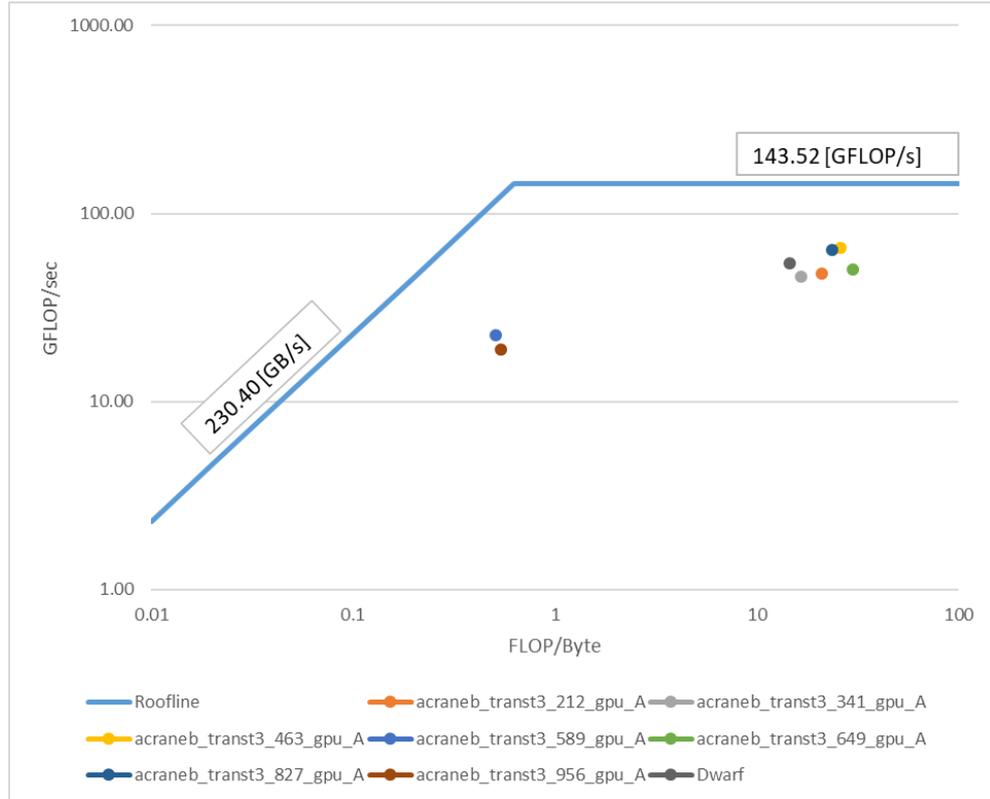

*Figure 4 Roofline for dwarf and individual kernels for GeForce 970 and size of domain 200x200x80*

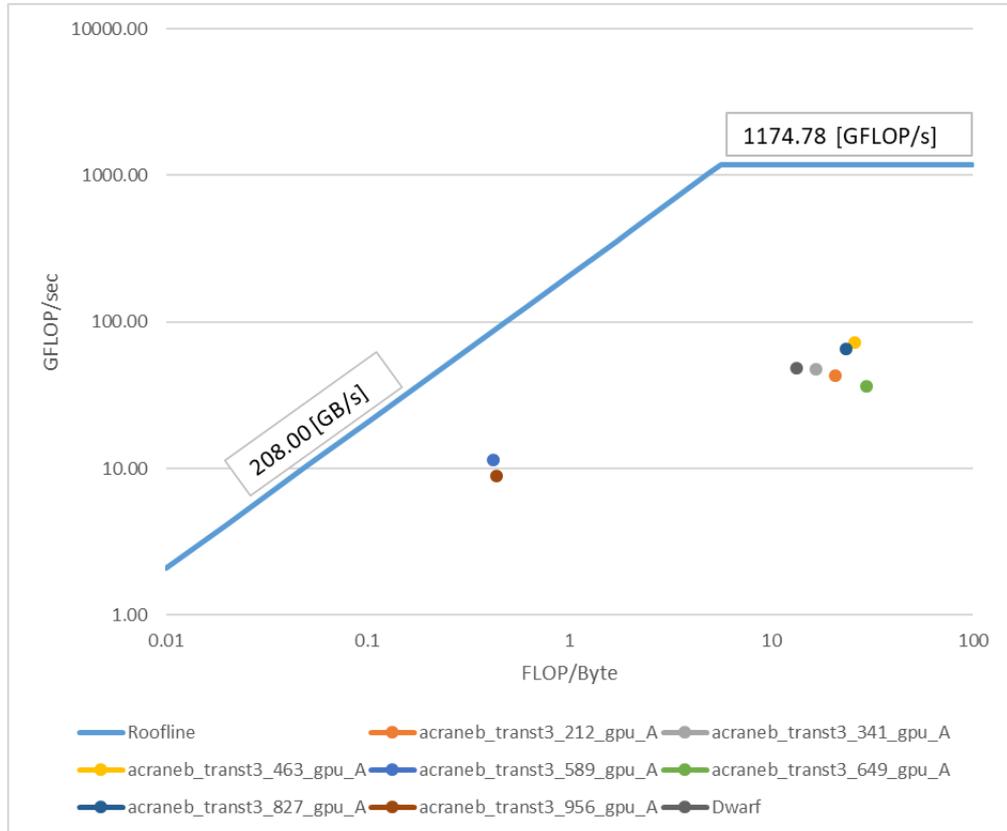

*Figure 5 Roofline for dwarf and individual kernels for Tesla K20m and size of domain 200x200x80*

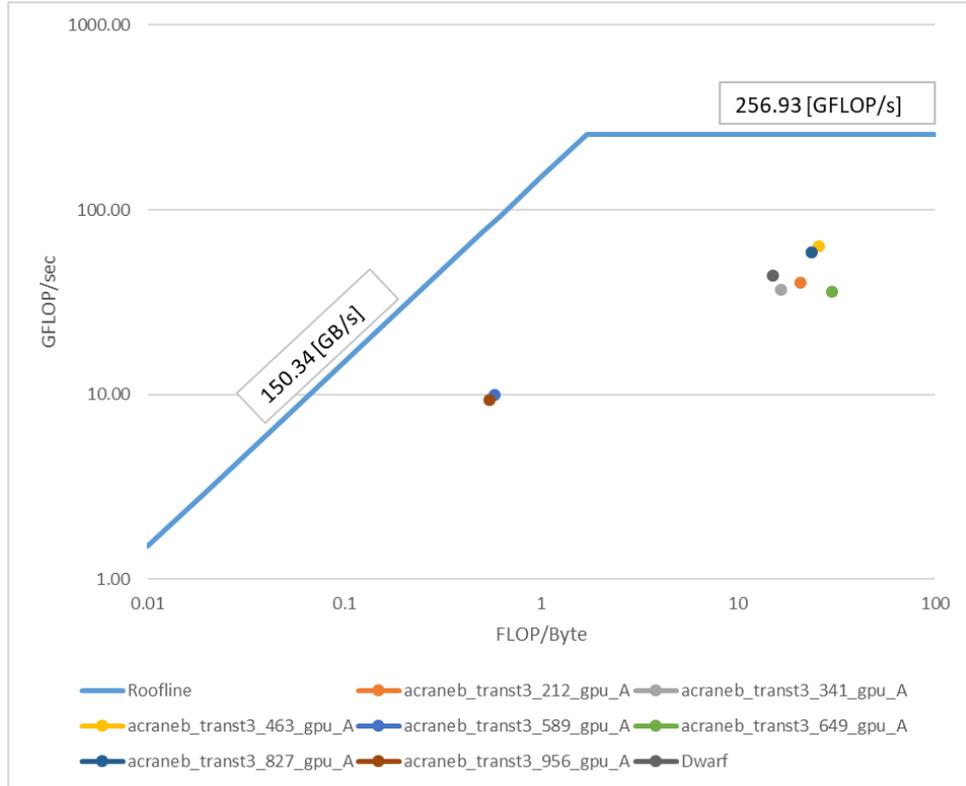

*Figure 6 Roofline for dwarf and individual kernels for Tesla 2070-q and size of domain 200x200x80*

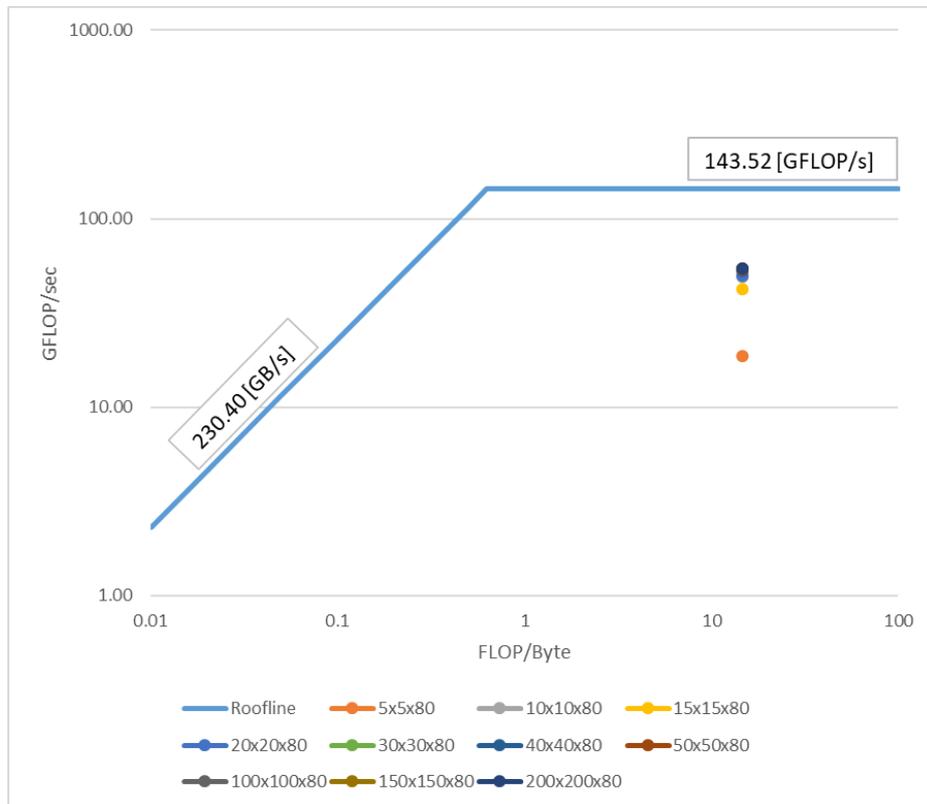

*Figure 7 Roofline for dwarf for GeForce 970 for different sizes of the domain*

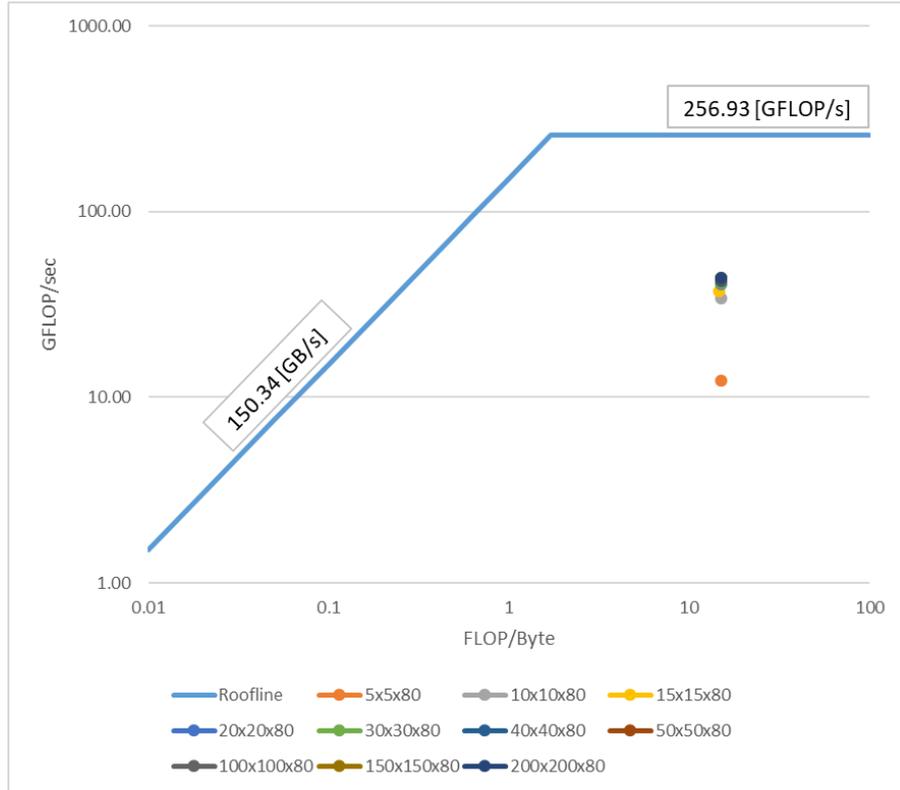

*Figure 8 Roofline for dwarf for Tesla K20m for different sizes of the domain*

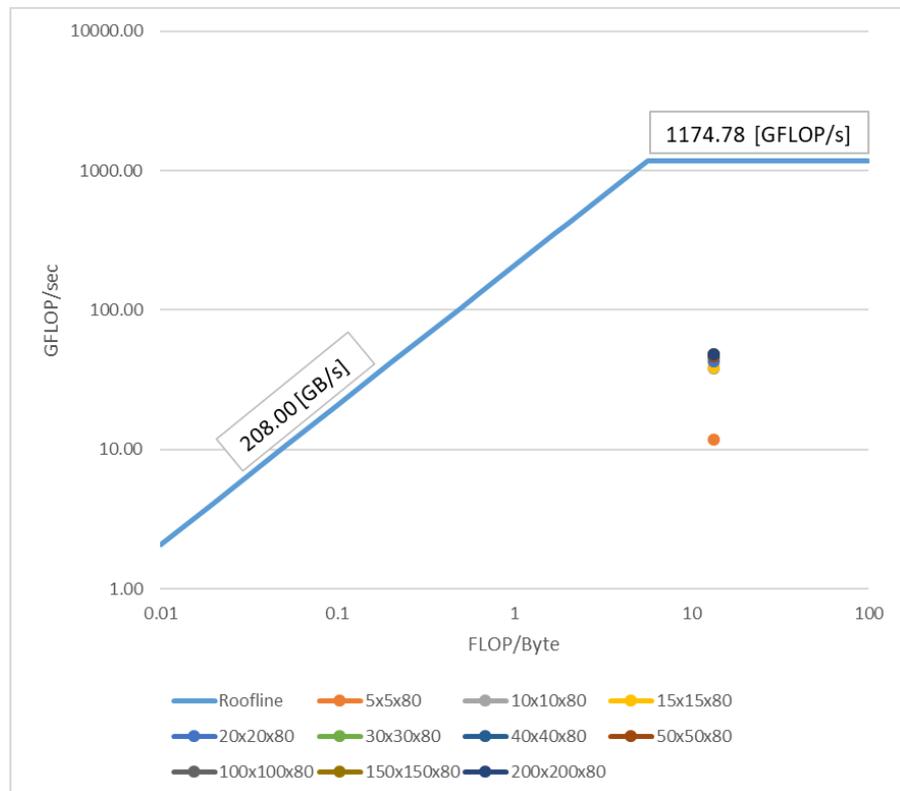

*Figure 9 Roofline for dwarf for Tesla 2070-q for different sizes of the domain*

| Kernel | W [Flop] | Q [Byte] | T [s] | W/Q | W/T [GFLOP/s] |
|---|---|---|---|---|---|
| acraneb_transt3_212_gpu_A | 8.85E+10 | 4.25E+09 | 1.78E+00 | 2.08E+01 | 4.99E+01 |
| acraneb_transt3_341_gpu_A | 7.05E+10 | 4.25E+09 | 1.44E+00 | 1.66E+01 | 4.89E+01 |
| acraneb_transt3_463_gpu_A | 1.63E+11 | 6.32E+09 | 2.52E+00 | 2.58E+01 | 6.48E+01 |
| acraneb_transt3_589_gpu_A | 4.67E+09 | 9.22E+09 | 2.44E-01 | 5.07E-01 | 1.91E+01 |
| acraneb_transt3_649_gpu_A | 1.59E+11 | 5.31E+09 | 3.23E+00 | 2.99E+01 | 4.92E+01 |
| acraneb_transt3_827_gpu_A | 1.49E+11 | 6.32E+09 | 2.39E+00 | 2.36E+01 | 6.25E+01 |
| acraneb_transt3_956_gpu_A | 4.44E+09 | 8.23E+09 | 2.51E-01 | 5.39E-01 | 1.77E+01 |
|  |  |  |  |  |  |
| SUM | 6.40E+11 | 4.39E+10 | 1.19E+01 | 1.46E+01 | 5.40E+01 |

*Table 1 Tabular form of roofline for dwarf and individual kernels for GeForce 970 and size of domain 200x200x80*

| Kernel | W [Flop] | Q [Byte] | T [s] | W/Q | W/T [GFLOP/s] |
|---|---|---|---|---|---|
| acraneb_transt3_212_gpu_A | 8.85E+10 | 4.25E+09 | 1.93E+00 | 2.08E+01 | 4.59E+01 |
| acraneb_transt3_341_gpu_A | 7.05E+10 | 4.25E+09 | 1.59E+00 | 1.66E+01 | 4.43E+01 |
| acraneb_transt3_463_gpu_A | 1.63E+11 | 6.32E+09 | 2.11E+00 | 2.58E+01 | 7.73E+01 |
| acraneb_transt3_589_gpu_A | 4.67E+09 | 1.12E+10 | 3.35E-01 | 4.19E-01 | 1.40E+01 |
| acraneb_transt3_649_gpu_A | 1.59E+11 | 5.31E+09 | 3.60E+00 | 2.99E+01 | 4.42E+01 |
| acraneb_transt3_827_gpu_A | 1.49E+11 | 6.32E+09 | 2.10E+00 | 2.36E+01 | 7.10E+01 |
| acraneb_transt3_956_gpu_A | 4.44E+09 | 1.02E+10 | 3.56E-01 | 4.37E-01 | 1.25E+01 |
|  |  |  |  |  |  |
| SUM | 6.40E+11 | 4.78E+10 | 1.20E+01 | 1.34E+01 | 5.32E+01 |

*Table 2 Tabular form of roofline for dwarf and individual kernels for Tesla K20m and size of domain 200x200x80*

| Kernel | W [Flop] | Q [Byte] | T [s] | W/Q | W/T [GFLOP/s] |
|---|---|---|---|---|---|
| acraneb_transt3_212_gpu_A | 8.85E+10 | 4.25E+09 | 1.83E+00 | 2.08E+01 | 4.85E+01 |
| acraneb_transt3_341_gpu_A | 7.05E+10 | 4.25E+09 | 1.55E+00 | 1.66E+01 | 4.55E+01 |
| acraneb_transt3_463_gpu_A | 1.63E+11 | 6.32E+09 | 2.34E+00 | 2.58E+01 | 6.98E+01 |
| acraneb_transt3_589_gpu_A | 4.67E+09 | 8.12E+09 | 3.20E-01 | 5.76E-01 | 1.46E+01 |
| acraneb_transt3_649_gpu_A | 1.59E+11 | 5.31E+09 | 3.12E+00 | 2.99E+01 | 5.09E+01 |
| acraneb_transt3_827_gpu_A | 1.49E+11 | 6.32E+09 | 2.27E+00 | 2.36E+01 | 6.58E+01 |
| acraneb_transt3_956_gpu_A | 4.44E+09 | 8.14E+09 | 3.22E-01 | 5.45E-01 | 1.38E+01 |
|  |  |  |  |  |  |
| SUM | 6.40E+11 | 4.27E+10 | 1.17E+01 | 1.50E+01 | 5.44E+01 |

*Table 3 Tabular form of roofline for dwarf and individual kernels for Tesla 2070-q and size of domain 200x200x80*

| Size | W [Flop] | Q [Byte] | T [s] | W/Q | W/T [GFLOP/s] |
|---|---|---|---|---|---|
| 200x200x80 | 6.40E+11 | 4.39E+10 | 1.18E+01 | 1.46E+01 | 5.43E+01 |
| 150x150x80 | 3.60E+11 | 2.47E+10 | 6.61E+00 | 1.46E+01 | 5.44E+01 |
| 100x100x80 | 1.60E+11 | 1.10E+10 | 2.94E+00 | 1.46E+01 | 5.43E+01 |
| 50x50x80 | 4.00E+10 | 2.74E+09 | 7.49E-01 | 1.46E+01 | 5.34E+01 |
| 40x40x80 | 2.56E+10 | 1.76E+09 | 4.84E-01 | 1.46E+01 | 5.29E+01 |
| 30x30x80 | 1.44E+10 | 9.88E+08 | 2.67E-01 | 1.46E+01 | 5.40E+01 |
| 20x20x80 | 6.40E+09 | 4.39E+08 | 1.29E-01 | 1.46E+01 | 4.97E+01 |
| 15x15x80 | 3.60E+09 | 2.47E+08 | 8.54E-02 | 1.46E+01 | 4.22E+01 |
| 10x10x80 | 1.60E+09 | 1.10E+08 | 3.25E-02 | 1.46E+01 | 4.91E+01 |
| 5x5x80 | 4.00E+08 | 2.74E+07 | 2.14E-02 | 1.46E+01 | 1.86E+01 |

*Table 4 Tabular form of roofline for dwarf for GeForce 970 for different sizes of the domain*

| Size | W [Flop] | Q [Byte] | T [s] | W/Q | W/T [GFLOP/s] |
|---|---|---|---|---|---|
| 200x200x80 | 6.40E+11 | 4.78E+10 | 1.33E+01 | 1.34E+01 | 4.81E+01 |
| 150x150x80 | 3.60E+11 | 2.69E+10 | 7.50E+00 | 1.34E+01 | 4.80E+01 |
| 100x100x80 | 1.60E+11 | 1.19E+10 | 3.34E+00 | 1.34E+01 | 4.79E+01 |
| 50x50x80 | 4.00E+10 | 2.99E+09 | 8.56E-01 | 1.34E+01 | 4.67E+01 |
| 40x40x80 | 2.56E+10 | 1.91E+09 | 5.51E-01 | 1.34E+01 | 4.64E+01 |
| 30x30x80 | 1.44E+10 | 1.07E+09 | 3.11E-01 | 1.34E+01 | 4.62E+01 |
| 20x20x80 | 6.40E+09 | 4.78E+08 | 1.49E-01 | 1.34E+01 | 4.29E+01 |
| 15x15x80 | 3.60E+09 | 2.69E+08 | 9.33E-02 | 1.34E+01 | 3.86E+01 |
| 10x10x80 | 1.60E+09 | 1.19E+08 | 4.20E-02 | 1.34E+01 | 3.81E+01 |
| 5x5x80 | 4.00E+08 | 2.99E+07 | 3.40E-02 | 1.34E+01 | 1.18E+01 |

*Table 5 Tabular form of roofline for dwarf for Tesla K20m for different sizes of the domain*

| Size | W [Flop] | Q [Byte] | T [s] | W/Q | W/T [GFLOP/s] |
|---|---|---|---|---|---|
| 200x200x80 | 6.40E+11 | 4.27E+10 | 1.46E+01 | 1.50E+01 | 4.39E+01 |
| 150x150x80 | 3.60E+11 | 2.40E+10 | 8.22E+00 | 1.50E+01 | 4.38E+01 |
| 100x100x80 | 1.60E+11 | 1.07E+10 | 3.66E+00 | 1.50E+01 | 4.37E+01 |
| 50x50x80 | 4.00E+10 | 2.67E+09 | 9.26E-01 | 1.50E+01 | 4.32E+01 |
| 40x40x80 | 2.56E+10 | 1.71E+09 | 6.05E-01 | 1.50E+01 | 4.23E+01 |
| 30x30x80 | 1.44E+10 | 9.61E+08 | 3.50E-01 | 1.50E+01 | 4.11E+01 |
| 20x20x80 | 6.40E+09 | 4.27E+08 | 1.58E-01 | 1.50E+01 | 4.05E+01 |
| 15x15x80 | 3.60E+09 | 2.46E+08 | 9.63E-02 | 1.46E+01 | 3.74E+01 |
| 10x10x80 | 1.60E+09 | 1.07E+08 | 4.69E-02 | 1.50E+01 | 3.41E+01 |
| 5x5x80 | 4.00E+08 | 2.67E+07 | 3.25E-02 | 1.50E+01 | 1.23E+01 |

*Table 6 Tabular form of roofline for dwarf for Tesla 2070-q for different sizes of the domain*

There are few interesting characteristics that can be observed:
- Kernels among each other differs in most of the counters, including time of execution, amount of work (W), amount of required data (Q) and performance.
- Amount of data required to perform kernels slightly differs between GPUs. It is important to note that this information was taken from the profiler – data requested by the kernel. The most probable reason of this behaviour is fact that OpenACC performed different optimisation schemes for different architectures.

- All kernels' executions have much lower performance comparing to the roofline. In our opinion it is caused by extensive synchronisation and inefficient data transfer, which result in long time of inactivity of the processing units.
- Surprisingly, the best performance was achieved by Nvidia GeForce 970, which has the lowest double precision performance among tested GPUs. We believe it is because this architecture can handle synchronisation and inefficient data transfer better.

Changing size of the domain has very limited impact on overall performance of the dwarf. Only the smallest domain was unable to provide enough parallelism and computations for the combination of tested architectures and benchmarked kernels. For that reason, we provided a simplification to the model assuming kernels have enough computations and parallelism to feed GPUs. To this end, we did not incorporate size of the domain into the DCworms simulator. It is worth noting that profiler and simulator have been simplified not to include data exchange between GPU and CPU as well as computational time spent on CPU.

### 3.4 Implementation

Taking into account ACRANABE2 dwarf and its kernels benchmarks, we propose model based on a set of counters provided by Nvidia CUDA profiler (see Table 7).

| Name of counter | Symbol | Description |
| --- | --- | --- |
| Duration(s) | $T$ | execution time of the kernel |
| Control Flow Instructions | $i_{cf}$ | Number of control flow instructions performed by all threads |
| Bit Convert Instructions | $i_{bc}$ | Number of bit convert instructions performed by all threads |
| Misc Instructions | $i_m$ | Number of miscellaneous instructions performed by all threads |
| Load/Store Instructions | $i_t$ | Number of load/store instructions performed by all threads |
| Integer Instructions | $i_i$ | Number of integer instructions performed by all threads |
| FP Instructions(Single) | $i_f$ | Number of floating point instructions performed by all threads |
| FP Instructions(Double) | $i_d$ | Number of double precision floating point instructions performed by all threads |
| Instructions Executed | $I_e$ | Instructions executed per warps |
| Active Cycles | $AC$ | Elapsed clock cycles summed from all SM |
| Executed IPC | $IPC_e$ | Instructions executed by all SM divided by active cycles i.e. how many instructions were executed in a single clock cycle; |
| Requested Global Load Throughput(bytes/sec) | $G_{rl}$ | Total size of data load requests by the kernel divided by the execution time of the kernel |
| Requested Global Store Throughput(bytes/sec) | $G_{rs}$ | Total size of data store requests by the kernel divided by the execution time of the kernel |
| Requested Non Coherent Global Load Throughput(bytes/sec) | $G_{rlnc}$ | Total size of data load requests from non-coherent memory by the kernel divided by the execution time of the kernel |
| Global Load Throughput(bytes/sec) | $G_l$ | Total size of data loaded by the kernel divided by the execution time of the kernel |
| Global Store Throughput(bytes/sec) | $G_s$ | Total size of data stored requests by the kernel divided by the execution time of the kernel |
| Non Coherent Global Memory Load Throughput(bytes/sec) | $G_{lnc}$ | Total size of data loaded from non-coherent memory by the kernel divided by the execution time of the kernel |
| Size of computational domain | $N$ | Product of all dimensions of computational domain; it is not the counter but it is used for further computations |

*Table 7 Nvidia CUDA profiler counters used in the model*

The performance model is based on kernel's achieved parallelism and efficiency of data load and store, in addition to amount of work (in specific numerical computations) that need to be performed. Each SM is executing instructions ($I_e$) which are intended to be executed in parallel by multiple threads, ideally by the whole warp (32 threads). As described in section 3.1, there are multiple conditions where only part of the warp can execute an instruction in parallel. The efficiency of the parallelism may be described as:

$$P_e = \frac{\sum_{x \in \{cf,m,t,f,d,bc,i\}} i_x}{I_e * 32}$$

We define $P_e$ as parallel thread execution efficiency. In the best situation this parameter is equal to 1. For similar reasons the multiprocessor may be unable to execute in an optimal way for particular architecture number of instructions in a single clock cycle. Many GPU architectures have the capability to execute multiple instructions in a single clock cycle per warp per SM and the actual number for an execution corresponds to the value of counter $IPC_e$.

The efficiency of data load and store is dependent on how the kernel is fulfilling the optimal patterns of accessing the data on a particular architecture. Its simplified version can be expressed by calculating percentage of data fetched by the chip that was actually requested by the kernel ($G_e$ – see Table 8).

Profiler counters are used to compute the characteristics of the kernels which are independent of the domain size. These variables are computed by dividing a counter value by the size of the domain (N). In order to obtain these characteristic one should also take the average of the kernel execution on different GPUs (the symbols containing in subscript word *avg*). The subscripts ending with the letter *n* are size independent. The counters taken to compute characteristics of kernels were measured for the largest size of the domain that fitted all benchmarked GPU architectures i.e. 200x200x80 (N = 3200000). The kernel specific characteristics as well as equations used for their computation are presented in Table 8. Kernel characteristics for dwarf ACRANEB2 are presented in Table 9.

| Symbol | Equation | Description |
| --- | --- | --- |
| $i_{avgn}$ | $i = \sum_{x \in \{cf,m,t,f,d,bc,i\}} i_x$ | Sum of all instructions performed per thread (averaged between architectures and normalized to the size of the domain) |
| $P_{iavg}$ | $P_i = 1 - \frac{i}{I_e * 32}$ | Percentage of time the kernel was inactive; |
| $P_{mavg}$ | $P_m = \frac{i_m}{I_e * 32}$ | Percentage of time the kernel was performing miscellaneous instructions |
| $P_{cflavg}$ | $P_{cfl} = \frac{i_{cfl}}{I_e * 32}$ | Percentage of time the kernel was performing control flow instructions |
| $G_{eavg}$ | $G_e = \frac{G_{rl} + G_{rs} + G_{rlnc}}{G_l + G_s + G_{lnc}}$ | Efficiency of global memory load and store, in other words, how much of the data fetched and stored was actually requested by the kernel |
| $G_{ravgn}$ | $G_e = (G_{rl} + G_{rs} + G_{rlnc}) * T$ | Total amount of data requested to be loaded and stored by the kernel |

| | |
|---|---|
| $i_{favgn}$, $i_{davgn}$, $i_{iavgn}$ | Arithmetic instructions averaged between different architectures and normalized to the size of domain |
| $I_{eavgn}$ | Instructions executed (counted in warps), averaged between architectures and normalized to the size of the domain |
| $IPC_{eavg}$ | |

Table 8 Kernel specific characteristics definition

| Symbol | kernel acraneb_transt3_ + | | | | | | |
|---|---|---|---|---|---|---|---|
| | 212_gpu_A | 341_gpu_A | 463_gpu_A | 589_gpu_A | 649_gpu_A | 827_gpu_A | 956_gpu_A |
| $P_{iavg}$ | 65.66% | 68.10% | 53.33% | 48.62% | 61.68% | 55.15% | 48.78% |
| $P_{mavg}$ | 8.34% | 7.67% | 11.88% | 6.82% | 9.47% | 11.41% | 7.01% |
| $P_{cflavg}$ | 1.43% | 1.33% | 1.88% | 2.04% | 1.58% | 1.81% | 2.05% |
| $G_{eavg}$ | 30.92% | 30.92% | 37.59% | 61.53% | 35.49% | 37.58% | 61.91% |
| $G_{ravgn}$ | 6.64E+02 | 6.64E+02 | 9.88E+02 | 1.48E+03 | 8.30E+02 | 9.88E+02 | 1.38E+03 |
| $i_{iavgn}$ | 3.73E+03 | 3.25E+03 | 5.53E+03 | 4.73E+03 | 5.40E+03 | 5.15E+03 | 4.72E+03 |
| $i_{favgn}$ | 9.17E+02 | 7.95E+02 | 1.23E+03 | 1.32E+02 | 1.71E+03 | 1.21E+03 | 1.32E+02 |
| $i_{davgn}$ | 9.08E+03 | 7.17E+03 | 1.68E+04 | 4.93E+02 | 1.62E+04 | 1.53E+04 | 4.68E+02 |
| $I_{eavgn}$ | 1.93E+03 | 1.69E+03 | 2.44E+03 | 4.25E+02 | 2.95E+03 | 2.32E+03 | 4.24E+02 |
| $IPC_{eavg}$ | 5.82E-01 | 6.35E-01 | 6.26E-01 | 7.27E-01 | 4.50E-01 | 6.07E-01 | 6.54E-01 |

Table 9 Kernel specific characteristics for dwarf ACRANEB2

| Description | Symbol | GPU | | |
|---|---|---|---|---|
| | | GeForce 970 | Tesla K20m | Tesla 2070-q |
| GPU Clock [Hz] | $C_G$ | 1.38E+09 | 7.06E+08 | 1.15E+09 |
| Memory Clock [Hz] | $C_M$ | 3.60E+09 | 2.60E+09 | 1.57E+09 |
| Memory Bus [bit] | $G_{bit}$ | 256 | 320 | 384 |
| Number of multiprocessors | SM | 13 | 13 | 14 |
| Number of double precision floating point operations that can be executed per clock cycle on SM | $O_{SMd}$ | 4 | 64 | 8 |
| Number of floating point operations that can be executed per clock cycle on SM | $O_{SMf}$ | 128 | 192 | 192 |
| Number of integer operations that can be executed per clock cycle on SM | $O_{SMi}$ | 128 | 160 | 160 |
| Peak Performance Double [FLOP/s] | $O_d$ | 1.44E+11 | 1.17E+12 | 2.57E+11 |
| Peak Performance Single [FLOP/s] | $O_f$ | 4.59E+12 | 3.52E+12 | 6.17E+12 |
| Peak Performance Integer [FLOP/s] | $O_i$ | 2.30E+12 | 1.47E+12 | 2.57E+12 |
| Memory Bandwidth [B/s] | $G_{max}$ | 2.30E+11 | 2.08E+11 | 1.50E+11 |

Table 10 Characteristics of GPUs

Table 10 presents characteristic of benchmarked GPUs that are utilised in the model. Table 11 presents the proposed GPU model.

| Symbol | Equation | Description |
|---|---|---|
| $T_i =$ | $\dfrac{P_{iavg} * I_{eavgn} * N}{IPC_{eavg} * C_G * SM}$ | Time of being inactive |
| $T_m =$ | $\dfrac{P_{mavg} * I_{eavgn} * N}{IPC_{eavg} * C_G * SM}$ | Time of performing miscellaneous instructions |
| $T_{cfl} =$ | $\dfrac{P_{cflavg} * I_{eavgn} * N}{IPC_{eavg} * C_G * SM}$ | Time of performing control flow instructions |
| $T_{cf} =$ | $\dfrac{N * i_{favgn} * 2}{O_f}$ | Time of performing floating point instructions; multiplication by two is caused by the fact that peak performance ($O_f$) is counted for multiply-add instructions (2 operations per instruction) and $I_{favgn}$ is representing instructions |
| $T_{cd} =$ | $\dfrac{N * i_{davgn} * 2}{O_d}$ | Time of performing double precision floating point instructions |
| $T_{ci} =$ | $\dfrac{N * i_{iavgn}}{O_i}$ | Time of performing integer instructions |
| $T_t =$ | $\dfrac{G_{ravgn} * N}{G_{max} * G_{eavg}}$ | Time to transfer data from/to the chip |
| $T_{sim} =$ | $\sum_{x \in \{i,m,cfl,t,cf,cd,ci\}} T_x$ | Time of the kernel execution from the simulator |

*Table 11 GPU model*

The model utilises precomputed kernel characteristics as well as properties of the GPUs in order to:
- Compute time spent for being inactive and perform miscellaneous and control flow instructions. These values are estimated from the average number of issued instructions on different architectures ($I_{eavgn}$), parallel efficiency ($P_{eavg}$) and instructions executed per clock cycle ($IPC_{eavg}$).
- Estimate time spent on different numerical computations on different data types. The model utilises the instruction counters ($i_{\{f,d,i\}avgn}$) and GPUs peak performance for particular data ($O_{\{f,d,i\}}$). Parallelism inefficiency is not taken into consideration in this step of the model since counters do not give enough information on parallelism for each type of executed instruction. Inactivity time already expresses parallel inefficiency of numerical computations;
- Estimate data load and store time. It is derived from the amount of data requested to be loaded and stored ($G_{ravgn}$), data transfer efficiency ($G_{eavg}$) for a particular kernel and peak bandwidth of the device ($G_{max}$). Moreover, we have observed that data transfer efficiency is lower for Fermi microarchitecture (Tesla 2070-q) and significantly lower when $G_{eagv} < 40\%$. This is why following adjustment has been proposed for this particular architecture:

$$G_{e(fermi)} = \begin{cases} \dfrac{1}{4} G_e \text{ for } G_e < 40\% \\ \dfrac{2}{3} G_e \text{ for } G_e \geq 40\% \end{cases}$$

The duration time estimated for each component of the model is then summed up resulting in total time of the execution. However, for optimised codes the transfer and computations should overlap, we did not observe this behaviour in case of OpenACC optimised kernels that is why the time is being summed up. The observations of behaviour of the model on tested GPUs indicate that probably some improvements would be required for newer microarchitectures. There is also visible inaccuracy of the model in some cases. The reason for that is that values of many counters, e.g. instruction counters other than related to numerical operations, instructions per clock cycle, differ between microarchitectures. These facts make it difficult to create consistent kernel characteristics. Nevertheless, we believe that further tuning could lead to overspecialisation for these particular benchmarked kernels and make the model unsuitable for the others.

### 3.5 Model verification

The verification was performed for individual kernels for the domain size 200x200x80 (Tables 12 - 14) as well as its quarter i.e. 100x100x80 (Tables 15 - 17). For other sizes only the cumulative time of dwarf execution was compared (Tables 18 - 20). The comparison is presented only in tabular form for clarity. Presented difference is the quotient of the simulated and profiling time minus one. Square error is computed on the values of difference. Figure 10 presents the difference between the simulation and profiling time over the size of the domain plotted for all GPUs. Figure 11 and 12 present both profiling and simulated time for benchmarked kernels and the same two sizes of domain.

|  | Time [s] | | | | | |
|---|---|---|---|---|---|---|
| **Kernel** | **Compute** | **Transfer** | **Inactive** | **Total** | **Simulated** | **Difference** |
| acraneb_transt3_212_gpu_A | 8.23E-01 | 5.96E-02 | 8.93E-01 | 1.78E+00 | 1.84E+00 | -3.38% |
| acraneb_transt3_341_gpu_A | 6.51E-01 | 5.96E-02 | 7.33E-01 | 1.44E+00 | 1.52E+00 | -4.90% |
| acraneb_transt3_463_gpu_A | 1.52E+00 | 7.30E-02 | 9.32E-01 | 2.52E+00 | 2.50E+00 | 0.99% |
| acraneb_transt3_589_gpu_A | 5.75E-02 | 6.70E-02 | 1.20E-01 | 2.44E-01 | 2.08E-01 | 17.59% |
| acraneb_transt3_649_gpu_A | 1.47E+00 | 6.50E-02 | 1.70E+00 | 3.23E+00 | 3.14E+00 | 2.86% |
| acraneb_transt3_827_gpu_A | 1.38E+00 | 7.30E-02 | 9.34E-01 | 2.39E+00 | 2.33E+00 | 2.25% |
| acraneb_transt3_956_gpu_A | 5.53E-02 | 6.20E-02 | 1.34E-01 | 2.51E-01 | 2.35E-01 | 6.66% |
| **Dwarf** | 5.95E+00 | 4.59E-01 | 5.44E+00 | 1.19E+01 | 1.18E+01 | 0.70% |

*Table 12 Verification for individual kernels; GeForce 970, domain size: 200x200x80, square error: 8%*

| | Time [s] | | | | | |
|---|---|---|---|---|---|---|
| **Kernel** | **Compute** | **Transfer** | **Inactive** | **Total** | **Simulated** | **Difference** |
| acraneb_transt3_212_gpu_A | 1.18E-01 | 6.61E-02 | 1.75E+00 | 1.93E+00 | 2.06E+00 | -6.31% |
| acraneb_transt3_341_gpu_A | 9.52E-02 | 6.61E-02 | 1.43E+00 | 1.59E+00 | 1.49E+00 | 6.74% |
| acraneb_transt3_463_gpu_A | 2.12E-01 | 8.09E-02 | 1.82E+00 | 2.11E+00 | 2.25E+00 | -6.00% |
| acraneb_transt3_589_gpu_A | 2.65E-02 | 7.42E-02 | 2.34E-01 | 3.35E-01 | 4.06E-01 | -17.60% |
| acraneb_transt3_649_gpu_A | 2.07E-01 | 7.20E-02 | 3.32E+00 | 3.60E+00 | 4.34E+00 | -17.03% |
| acraneb_transt3_827_gpu_A | 1.93E-01 | 8.09E-02 | 1.83E+00 | 2.10E+00 | 2.27E+00 | -7.39% |
| acraneb_transt3_956_gpu_A | 2.61E-02 | 6.87E-02 | 2.61E-01 | 3.56E-01 | 4.99E-01 | -28.54% |
| | | | | | | |
| **Dwarf** | **8.78E-01** | **5.09E-01** | **1.06E+01** | **1.20E+01** | **1.33E+01** | **-9.65%** |

*Table 13 Verification for individual kernels; Tesla K20m, domain size: 200x200x80, square error: 15%*

| | Time [s] | | | | | |
|---|---|---|---|---|---|---|
| **Kernel** | **Compute** | **Transfer** | **Inactive** | **Total** | **Simulated** | **Difference** |
| acraneb_transt3_212_gpu_A | 4.63E-01 | 3.66E-01 | 9.98E-01 | 1.83E+00 | 2.19E+00 | -16.77% |
| acraneb_transt3_341_gpu_A | 3.67E-01 | 3.66E-01 | 8.19E-01 | 1.55E+00 | 1.90E+00 | -18.48% |
| acraneb_transt3_463_gpu_A | 8.53E-01 | 4.48E-01 | 1.04E+00 | 2.34E+00 | 2.57E+00 | -8.96% |
| acraneb_transt3_589_gpu_A | 3.66E-02 | 1.49E-01 | 1.34E-01 | 3.20E-01 | 4.71E-01 | -32.13% |
| acraneb_transt3_649_gpu_A | 8.25E-01 | 3.98E-01 | 1.90E+00 | 3.12E+00 | 4.43E+00 | -29.58% |
| acraneb_transt3_827_gpu_A | 7.76E-01 | 4.48E-01 | 1.04E+00 | 2.27E+00 | 2.54E+00 | -10.69% |
| acraneb_transt3_956_gpu_A | 3.54E-02 | 1.37E-01 | 1.49E-01 | 3.22E-01 | 4.72E-01 | -31.78% |
| | | | | | | |
| **Dwarf** | **3.36E+00** | **2.31E+00** | **6.08E+00** | **1.17E+01** | **1.46E+01** | **-19.43%** |

*Table 14 Verification for individual kernels; Tesla 2070-q, domain size: 200x200x80, square error: 23%*

| | Time [s] | | | | | |
|---|---|---|---|---|---|---|
| **Kernel** | **Compute** | **Transfer** | **Inactive** | **Total** | **Simulated** | **Difference** |
| acraneb_transt3_212_gpu_A | 2.06E-01 | 1.49E-02 | 2.23E-01 | 4.44E-01 | 4.60E-01 | -3.62% |
| acraneb_transt3_341_gpu_A | 1.63E-01 | 1.49E-02 | 1.83E-01 | 3.61E-01 | 3.79E-01 | -4.82% |
| acraneb_transt3_463_gpu_A | 3.79E-01 | 1.83E-02 | 2.33E-01 | 6.31E-01 | 6.32E-01 | -0.23% |
| acraneb_transt3_589_gpu_A | 1.44E-02 | 1.67E-02 | 3.00E-02 | 6.11E-02 | 5.24E-02 | 16.63% |
| acraneb_transt3_649_gpu_A | 3.67E-01 | 1.62E-02 | 4.24E-01 | 8.07E-01 | 7.76E-01 | 4.09% |
| acraneb_transt3_827_gpu_A | 3.45E-01 | 1.83E-02 | 2.34E-01 | 5.97E-01 | 5.86E-01 | 1.79% |
| acraneb_transt3_956_gpu_A | 1.38E-02 | 1.55E-02 | 3.34E-02 | 6.28E-02 | 5.91E-02 | 6.13% |
| | | | | | | |
| **Dwarf** | **1.49E+00** | **1.15E-01** | **1.36E+00** | **2.96E+00** | **2.94E+00** | **0.62%** |

*Table 15 Verification for individual kernels; GeForce 970, domain size: 100x100x80, square error: 7%*

| Kernel | Time [s] | | | | | |
|---|---|---|---|---|---|---|
| | Compute | Transfer | Inactive | Total | Simulated | Difference |
| acraneb_transt3_212_gpu_A | 2.96E-02 | 1.65E-02 | 4.36E-01 | 4.82E-01 | 5.18E-01 | -6.89% |
| acraneb_transt3_341_gpu_A | 2.38E-02 | 1.65E-02 | 3.58E-01 | 3.98E-01 | 3.74E-01 | 6.46% |
| acraneb_transt3_463_gpu_A | 5.29E-02 | 2.02E-02 | 4.55E-01 | 5.28E-01 | 5.65E-01 | -6.43% |
| acraneb_transt3_589_gpu_A | 6.61E-03 | 1.85E-02 | 5.85E-02 | 8.37E-02 | 1.02E-01 | -17.74% |
| acraneb_transt3_649_gpu_A | 5.16E-02 | 1.80E-02 | 8.30E-01 | 8.99E-01 | 1.09E+00 | -17.35% |
| acraneb_transt3_827_gpu_A | 4.83E-02 | 2.02E-02 | 4.57E-01 | 5.25E-01 | 5.69E-01 | -7.74% |
| acraneb_transt3_956_gpu_A | 6.53E-03 | 1.72E-02 | 6.54E-02 | 8.91E-02 | 1.25E-01 | -28.91% |
| **Dwarf** | **2.19E-01** | **1.27E-01** | **2.66E+00** | **3.01E+00** | **3.34E+00** | **-10.02%** |

*Table 16 Verification for individual kernels; Tesla K20m, domain size: 100x100x80, square error: 15%*

| Kernel | Time [s] | | | | | |
|---|---|---|---|---|---|---|
| | Compute | Transfer | Inactive | Total | Simulated | Difference |
| acraneb_transt3_212_gpu_A | 1.16E-01 | 9.14E-02 | 2.49E-01 | 4.57E-01 | 5.50E-01 | -16.97% |
| acraneb_transt3_341_gpu_A | 9.18E-02 | 9.14E-02 | 2.05E-01 | 3.88E-01 | 4.77E-01 | -18.65% |
| acraneb_transt3_463_gpu_A | 2.13E-01 | 1.12E-01 | 2.60E-01 | 5.85E-01 | 6.44E-01 | -9.12% |
| acraneb_transt3_589_gpu_A | 9.16E-03 | 3.73E-02 | 3.35E-02 | 8.00E-02 | 1.18E-01 | -32.45% |
| acraneb_transt3_649_gpu_A | 2.06E-01 | 9.96E-02 | 4.74E-01 | 7.80E-01 | 1.11E+00 | -29.82% |
| acraneb_transt3_827_gpu_A | 1.94E-01 | 1.12E-01 | 2.61E-01 | 5.67E-01 | 6.36E-01 | -10.84% |
| acraneb_transt3_956_gpu_A | 8.84E-03 | 3.43E-02 | 3.74E-02 | 8.05E-02 | 1.19E-01 | -32.06% |
| **Dwarf** | **8.39E-01** | **5.78E-01** | **1.52E+00** | **2.94E+00** | **3.66E+00** | **-19.64%** |

*Table 17 Verification for individual kernels; Tesla 2070-q, domain size: 100x100x80, square error: 23%*

| Size | W [Flop] | Q [Byte] | T sim [s] | W/Q | W/T [GFLOP/s] | T prof [s] | Difference |
|---|---|---|---|---|---|---|---|
| 5x5x80 | 4.00E+08 | 2.74E+07 | 7.41E-03 | 1.46E+01 | 5.40E+01 | 2.14E-02 | -65.46% |
| 10x10x80 | 1.60E+09 | 1.10E+08 | 2.96E-02 | 1.46E+01 | 5.40E+01 | 3.25E-02 | -8.95% |
| 15x15x80 | 3.60E+09 | 2.47E+08 | 6.67E-02 | 1.46E+01 | 5.40E+01 | 8.54E-02 | -21.89% |
| 20x20x80 | 6.40E+09 | 4.39E+08 | 1.19E-01 | 1.46E+01 | 5.40E+01 | 1.29E-01 | -7.97% |
| 30x30x80 | 1.44E+10 | 9.88E+08 | 2.67E-01 | 1.46E+01 | 5.40E+01 | 2.69E-01 | -0.87% |
| 40x40x80 | 2.56E+10 | 1.76E+09 | 4.74E-01 | 1.46E+01 | 5.40E+01 | 4.84E-01 | -2.04% |
| 50x50x80 | 4.00E+10 | 2.74E+09 | 7.41E-01 | 1.46E+01 | 5.40E+01 | 7.49E-01 | -1.10% |
| 100x100x80 | 1.60E+11 | 1.10E+10 | 2.96E+00 | 1.46E+01 | 5.40E+01 | 2.94E+00 | 0.62% |
| 150x150x80 | 3.60E+11 | 2.47E+10 | 6.67E+00 | 1.46E+01 | 5.40E+01 | 6.61E+00 | 0.83% |
| 200x200x80 | 6.40E+11 | 4.39E+10 | 1.19E+01 | 1.46E+01 | 5.40E+01 | 1.18E+01 | 0.70% |

*Table 18 Verification the results of simulation for all sizes, for the whole Dwarf, GeForce 970*

| Size | W [Flop] | Q [Byte] | T sim [s] | W/Q | W/T [GFLOP/s] | T prof. [s] | Difference |
|---|---|---|---|---|---|---|---|
| 5x5x80 | 4.00E+08 | 2.99E+07 | 7.52E-03 | 1.34E+01 | 5.32E+01 | 3.40E-02 | -77.89% |
| 10x10x80 | 1.60E+09 | 1.19E+08 | 3.01E-02 | 1.34E+01 | 5.32E+01 | 4.20E-02 | -28.45% |
| 15x15x80 | 3.60E+09 | 2.69E+08 | 6.76E-02 | 1.34E+01 | 5.32E+01 | 9.33E-02 | -27.49% |
| 20x20x80 | 6.40E+09 | 4.78E+08 | 1.20E-01 | 1.34E+01 | 5.32E+01 | 1.49E-01 | -19.30% |
| 30x30x80 | 1.44E+10 | 1.07E+09 | 2.71E-01 | 1.34E+01 | 5.32E+01 | 3.11E-01 | -13.08% |
| 40x40x80 | 2.56E+10 | 1.91E+09 | 4.81E-01 | 1.34E+01 | 5.32E+01 | 5.51E-01 | -12.67% |
| 50x50x80 | 4.00E+10 | 2.99E+09 | 7.52E-01 | 1.34E+01 | 5.32E+01 | 8.56E-01 | -12.21% |
| 100x100x80 | 1.60E+11 | 1.19E+10 | 3.01E+00 | 1.34E+01 | 5.32E+01 | 3.34E+00 | -10.02% |
| 150x150x80 | 3.60E+11 | 2.69E+10 | 6.76E+00 | 1.34E+01 | 5.32E+01 | 7.50E+00 | -9.76% |
| 200x200x80 | 6.40E+11 | 4.78E+10 | 1.20E+01 | 1.34E+01 | 5.32E+01 | 1.33E+01 | -9.65% |

*Table 19 Verification the results of simulation for all sizes, for the whole Dwarf, Tesla K20m*

| Size | W [Flop] | Q [Byte] | T sim [s] | W/Q | W/T [GFLOP/s] | T prof. [s] | Difference |
|---|---|---|---|---|---|---|---|
| 5x5x80 | 4.00E+08 | 2.67E+07 | 7.34E-03 | 1.50E+01 | 5.44E+01 | 3.25E-02 | -77.40% |
| 10x10x80 | 1.60E+09 | 1.07E+08 | 2.94E-02 | 1.50E+01 | 5.44E+01 | 4.69E-02 | -37.35% |
| 15x15x80 | 3.60E+09 | 2.46E+08 | 6.61E-02 | 1.46E+01 | 5.44E+01 | 9.63E-02 | -31.37% |
| 20x20x80 | 6.40E+09 | 4.27E+08 | 1.17E-01 | 1.50E+01 | 5.44E+01 | 1.58E-01 | -25.65% |
| 30x30x80 | 1.44E+10 | 9.61E+08 | 2.64E-01 | 1.50E+01 | 5.44E+01 | 3.50E-01 | -24.58% |
| 40x40x80 | 2.56E+10 | 1.71E+09 | 4.70E-01 | 1.50E+01 | 5.44E+01 | 6.05E-01 | -22.31% |
| 50x50x80 | 4.00E+10 | 2.67E+09 | 7.34E-01 | 1.50E+01 | 5.44E+01 | 9.26E-01 | -20.74% |
| 100x100x80 | 1.60E+11 | 1.07E+10 | 2.94E+00 | 1.50E+01 | 5.44E+01 | 3.66E+00 | -19.64% |
| 150x150x80 | 3.60E+11 | 2.40E+10 | 6.61E+00 | 1.50E+01 | 5.44E+01 | 8.22E+00 | -19.63% |
| 200x200x80 | 6.40E+11 | 4.27E+10 | 1.17E+01 | 1.50E+01 | 5.44E+01 | 1.46E+01 | -19.43% |

*Table 20 Verification the results of simulation for all sizes, for the whole Dwarf, Tesla 2070-q*

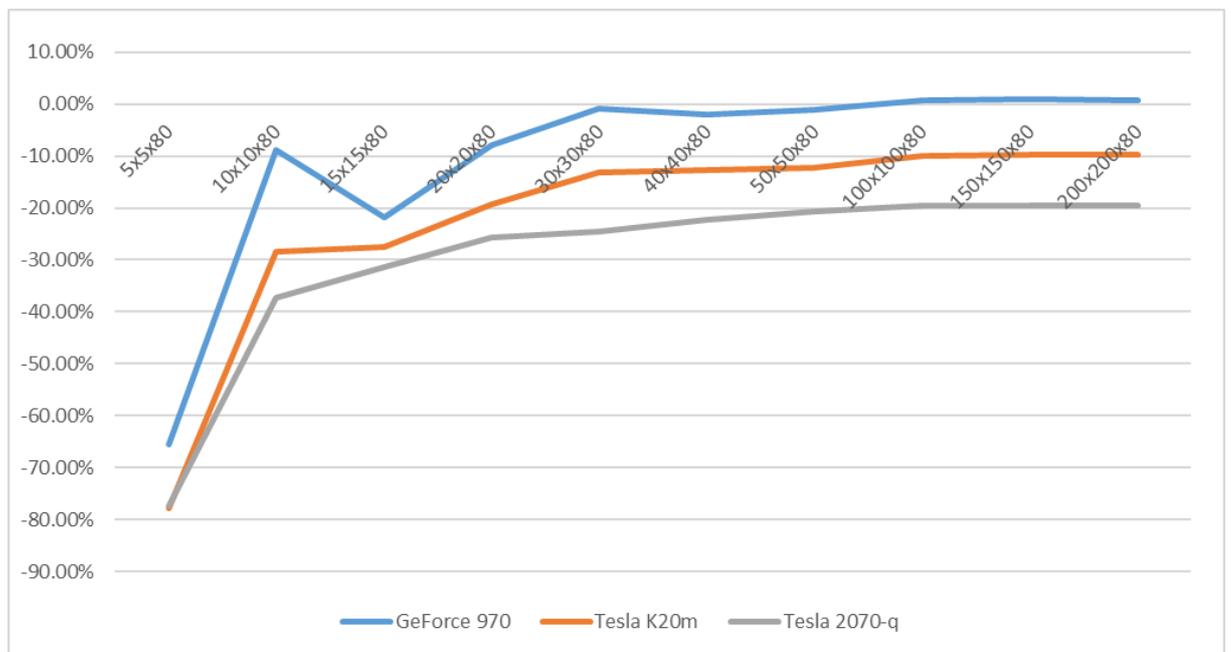

*Figure 10 The difference between simulation and profiling time for the whole dwarf and all benchmarked sizes*

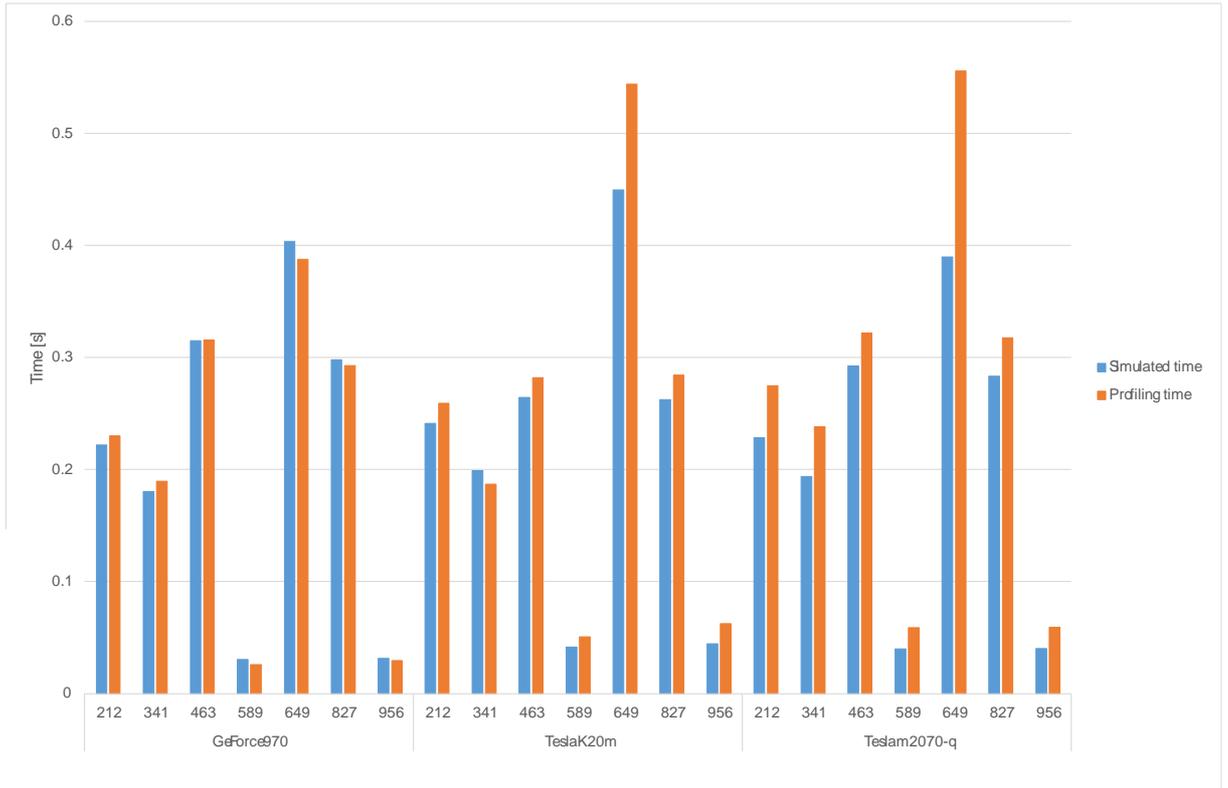

*Figure 11 Comparison of simulated and profiling time for all GPUs for all kernels for size of the domain equal to 100x100x80;*

*for clarity kernels have been abbreviated to differentiating part of their names*

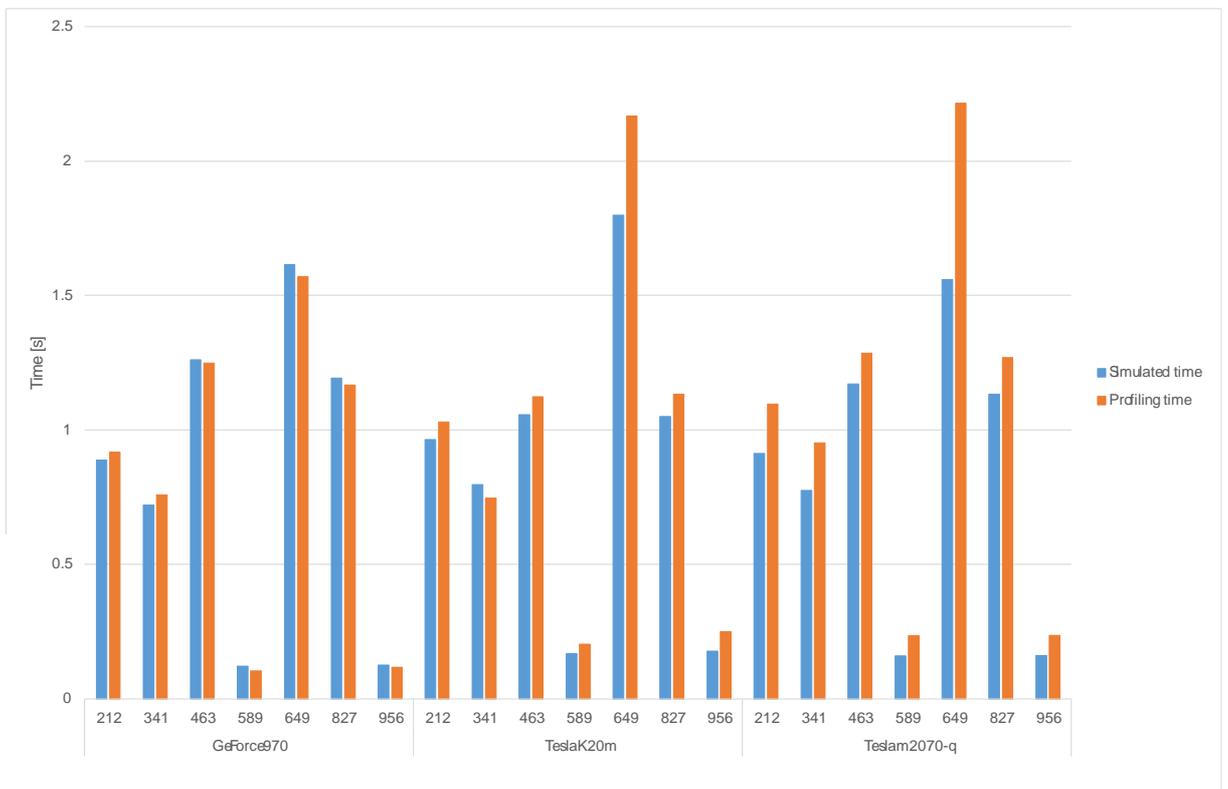

*Figure 12 Comparison of simulated and profiling time for all GPUs for all kernels for size of the domain equal to 200x200x80;*

*for clarity kernels have been abbreviated to differentiating part of their names.*

The verification processed proved almost perfect estimation of execution for GeForce 970 and sufficient for Tesla K20m. Fermi microarchitecture have significant differences comparing to the two other architectures and thus the model has worse estimation of performance for this GPU device. Further work and tuning should be applied in order to improve performance for that particular architecture. Nevertheless it may occur not worth the hassle, since Fermi is discontinued architecture and there is a significant chance that fine-tuning of the model for particular characteristics of the microarchitectures may result in its overspecialization and makes it useless for different computational problems and upcoming microarchitectures.

The influence of size of the domain is significant but only for very small sizes, thus it can be neglected.

As described in this deliverable, GPU architecture has a very complex nature. The number of patterns to be followed in order to achieve optimal performance as well as the number of varying counters and characteristics makes it very difficult to estimate time of kernel execution. In our opinion proposed model and DCWorms simulator proved its value in aforementioned circumstances.

## 4 Model for multinode

### 4.1 State-of-the-art

The performance models for the modern heterogeneous processing units that incorporates sophisticated memory hierarchies should be simple and efficient in order to explore properties of recent hardware units. The Roofline model allows to analyse, model and predict application performance based on a processing unit's computation and communication capabilities (Williams, Waterman i Patterson 2009). The application is modelled as a ratio of arithmetic operations to number of bytes sent through the memory hierarchy. The performance of a simple von Neumann architecture that contains two levels of memory hierarchy can be predicted with the Roofline model. The model can be extended to support a more complex memory hierarchy with a multi-level caches (Treibig and Hager 2010). Recently, it has been extended to model the energy consumption of GPUs (Choi and others 2013). The authors assumed that time per work (arithmetic) operation and time per memory operation are estimated with the hardware peak throughput values, whereas the energy cost is estimated using a linear regression based on real experiments.

For the multinode configuration, the Roofline model is not enough since it does not capture metrics such as time-to-solution or energy-to-solution. Another issue concerns communication pattern between nodes. In general, adding more nodes results in shrinking computational domain, thus less work for each node is required which results in increased computational efficiency. On the other hand, the overall amount of data needed to be exchanged between nodes increases, which may lead to situation where communication is more costly in comparison to computations. The most important optimisation recommendation would be to limit communication between nodes as much as possible, or to hide communication by computations, which however may require significant changes to the algorithms.

### 4.2 Spherical harmonics

The Spherical Harmonics dwarf is used as a building block in different weather models such as Integrated Forecast System (IFS) or ARPEGE (Action de Recherche Petite Echelle Grande Echelle). The dwarf implements spectral transform on a sphere where a Fast Fourier Transform (FFT) is applied in longitude and a Legendre transform (LT) in latitude, see Figure 13. The computational complexity of FFT is ≈ $O(N \log N)$, where $N$ is the cut off spectral truncation wavenumber. The Legendre transform has computational complexity of $O(N^2)$ which increases with horizontal resolution. Details on Spherical Harmonics dwarf can be found in Deliverable D1.1 (Section 4.1).

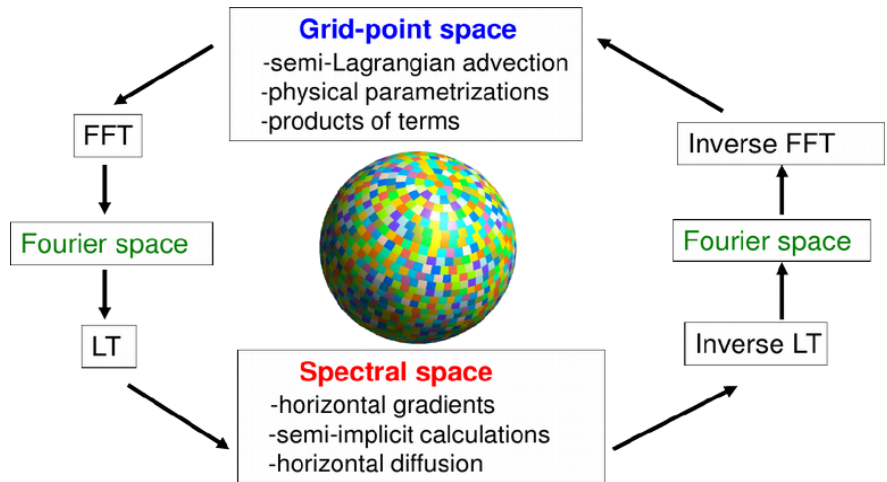

*Figure 13: Spherical Harmonics dwarf*

### 4.3 Implementation and usage

| Mnemonic name | HSW |
|---|---|
| CPU Model | Xeon E5-2697 v3 |
| Microarch. | Haswell EP |
| Process size | 22 nm |
| Release date | Q3 2014 |
| ISA | AVX2 |
| Cores | 14 |
| Sockets | 2 |
| HT | off |
| Frequency [GHz] | 2.6 |
| L1/core | 32KB |
| L2/core | 256KB |
| L3 | 35MB |
| DDR4 | 2133 MHz |
| Performance DP/core [GFLOP/s] | 20.8 |

*Table 21 Intel Xeon E5-2697v3 specification*

The performance model for CPU multinode has been provided on Spherical Harmonics dwarf example. In order to model performance on a multinode system, memory bandwidth of the single CPU, processing unit characteristic and amount of dwarf *W* (amount of work) and *Q* (amount of exchanged data within memory) knowledge is required. This is done by estimating number of dwarf iterations, number of bytes and FLOP per iteration, number of FLOP executed, execution time per FLOP and per byte moved, and number of all bytes moved through the memory hierarchy including L1, L2, L3 caches and DRAM. These parameters are collected using Intel Advisor tool and analysis of loops that are executed for more that 2% of total execution time. The number of FLOPS is estimated by using following equation:

```
trans nsmax2 * grid n * 0.833 * dwarf iterations * dwarf fields * Flop Per
Iterations
```

For example, the *ledir->dgemm_315* loop of SH dwarf for TCO639 testcase (see Table 22 for input parameters) has number of FLOPs equalled to:

$$639^2 * 640 * 0.833 * 100 * 200 * 2 = 8.70736\text{E}+12$$

|        | Resolution [km] | GRIDN | PTS     | NSMAX | ITER | F   |
|--------|-----------------|-------|---------|-------|------|-----|
| TL159  | 125             | 80    | 35718   | 159   | 100  | 274 |
| TCO639 | 16              | 640   | 1661440 | 639   | 100  | 200 |

*Table 22 Spherical Harmonics input variables values for different test cases*

The number of bytes moved is estimated using following formula:

```
number of FLOP = trans nsmax² * grid n * 0.833 * dwarf iterations * dwarf
fields * Bytes Per Iterations
```

or based on arithmetic intensity:

```
bytes moved = number of FLOP / AI
```

which for aforementioned *ledir->dgemm_315* loop gives us (see Table 23 for input parameters):

```
bytes moved = 8.70736E+12 / 0.083 = 1.04509E+14
```

| Loop | Type | Memory, GB | Memory, Bytes Per Iteration | AI |
|---|---|---|---|---|
| [loop in dgemm at dgemm.f:315] | Vectorized (Body; Peeled; Remainder) | 1108929 | 88.789 | 0.083 |
| [child]-[loop in dgemm at dgemm.f:315] | Vectorized (Body) | 1078909 | 96 | 0.083 |
| [child]-[loop in dgemm at dgemm.f:315] | Peeled | 14209 | 24 | 0.083 |
| [child]-[loop in dgemm at dgemm.f:315] | Remainder | 15811 | 24 | 0.083 |

*Table 23 Memory for the dgemm loop called from the leinv function*

Table 24 presents calculated values *W* and *Q* for considered dwarf. Only 3 loops employ some computations, whereas other 12 move data between different structures within memory.

| SH TCO639 loop | W Per iter. | W [FLOP] | I = W/Q | Q per iter. | Q [Bytes] |
|---|---|---|---|---|---|
| ledir->dgemm_327 | 2 | NSMAX² * GRIDN * 0.833 * ITER * F * W Per Iter.= 8.70736E+12 | 0.187 | 10.69 | W/I = 4.65232E+13 |
| leinv->dgemm_315 | 2 | NSMAX² * GRIDN * 0.833 * ITER * F * W Per Iter.= 8.70736E+12 | 0.083 | 24 | W/I = 1.04509E+14 |
| Trmtol_134 | - | - | - | 16.00 | PTS * ITER * F * Q Per Iter. = 5.31661E+11 |
| Trltom_130 | - | - | - | 16.00 | PTS * ITER * F * Q Per Iter. = 5.31661E+11 |
| ftinv_ctl->fourier_in_54 | - | - | - | 16.00 | PTS * ITER * F * Q Per Iter. = 5.31661E+11 |
| prfi1b_91 | - | - | - | 16.00 | PTS * ITER * F * Q Per Iter. = 5.31661E+11 |
| ltdir_ctl->updsp_132 | - | - | - | 11.398 | PTS * ITER * F * Q Per Iter. = 3.78742E+11 |

| asre1b_88 | 1 | PTS * ITER * F * W Per Iter. = 33228800000 | - | 16.00 | PTS * ITER * F * Q Per Iter. = 5.31661E+11 |
|---|---|---|---|---|---|
| Trgtol_434 | - | - | - | 22.00 | PTS * ITER * F * Q Per Iter. = 7.31034E+11 |
| Trltog_433 | - | - | - | 20.00 | PTS * ITER * F * Q Per Iter. = 6.64576E+11 |
| ftdir_104 | - | - | - | 16.00 | PTS * ITER * F * Q Per Iter. = 5.31661E+11 |
| prfi2b_80 | - | - | - | 16.00 | PTS * ITER * F * Q Per Iter. = 5.31661E+11 |
| fourier_out_53 | - | - | - | 16.00 | PTS * ITER * F * Q Per Iter. = 5.31661E+11 |
| Leinv_179 | - | - | - | 10.82 | PTS * ITER * F * Q Per Iter. = 3.59536E+11 |
| Leinv_142 | - | - | - | 10.82 | PTS * ITER * F * Q Per Iter. = 3.59536E+11 |

*Table 24 W and Q values for the performance model*

The next step is to correlate processing unit, CPU in our case, with the loops. For the compute part, linear programming is used to find U coefficient that meets following equation:

$$P(f,n) * U = M(f,n)$$

where *P* is the performance of a CPU and *M* is the performance achieved by loop, for a given number of CPU cores *n* and its frequency *f*. The performance of a processing unit is calculated using following equation:

```
frequency * vector size * number of vector operations per clock * number of cores utilised
```

Table 25 presents CPU performance for our case, i.e. Intel Xeon E5-2697v3.

| Freq. | 1.2 | 1.4 | 1.6 | 1.8 | 2 | 2.4 | 2.6 | trubo |
|---|---|---|---|---|---|---|---|---|
| Threads | GFLOP/s | | | | | | | |
| 1 | 9.600 | 11.200 | 12.800 | 14.400 | 16.000 | 19.200 | 20.800 | 28.800 |
| 2 | 19.200 | 22.400 | 25.600 | 28.800 | 32.000 | 38.400 | 41.600 | 57.600 |
| 3 | 28.800 | 33.600 | 38.400 | 43.200 | 48.000 | 57.600 | 62.400 | 81.600 |
| 4 | 38.400 | 44.800 | 51.200 | 57.600 | 64.000 | 76.800 | 83.200 | 108.800 |
| 5 | 48.000 | 56.000 | 64.000 | 72.000 | 80.000 | 96.000 | 104.000 | 128.000 |
| 6 | 57.600 | 67.200 | 76.800 | 86.400 | 96.000 | 115.200 | 124.800 | 153.600 |
| 7 | 67.200 | 78.400 | 89.600 | 100.800 | 112.000 | 134.400 | 145.600 | 179.200 |
| 8 | 76.800 | 89.600 | 102.400 | 115.200 | 128.000 | 153.600 | 166.400 | 204.800 |
| 9 | 86.400 | 100.800 | 115.200 | 129.600 | 144.000 | 172.800 | 187.200 | 230.400 |
| 10 | 96.000 | 112.000 | 128.000 | 144.000 | 160.000 | 192.000 | 208.000 | 256.000 |
| 11 | 105.600 | 123.200 | 140.800 | 158.400 | 176.000 | 211.200 | 228.800 | 281.600 |
| 12 | 115.200 | 134.400 | 153.600 | 172.800 | 192.000 | 230.400 | 249.600 | 307.200 |
| 13 | 124.800 | 145.600 | 166.400 | 187.200 | 208.000 | 249.600 | 270.400 | 332.800 |
| 14 | 134.400 | 156.800 | 179.200 | 201.600 | 224.000 | 268.800 | 291.200 | 358.400 |

*Table 25 Performance of Intel Xeon E5-2697v3*

*Frequency is given in GHz, bandwidth in GB/s*

To model data movement, linear programming is used to find coefficients *V*, *X*, *Y* and *Z* are found that meet following equation:

$$L1(f,n) * V + L2(f,n) * X + L3(f,n) * Y + DRAM(f,n) * Z = R(f,n)$$

where *f* is the frequency of CPU, *n* is the number of used cores, *R* is the measured bandwidth of a loop (that runs on the (*f*,*n*) state of the CPU). L1, L2, L3 and DRAM are measured memory bandwidth for different CPU frequencies and number of cores, which are presented in Table 26, Table 27, Table 28 and Table 29 respectively.

| Freq. | 1.2 | 1.4 | 1.6 | 1.8 | 2 | 2.4 | 2.6 | turbo |
|---|---|---|---|---|---|---|---|---|
| Threads | | | | L1 | | | | |
| 1 | 57.659 | 67.204 | 76.981 | 86.420 | 96.135 | 115.523 | 125.990 | 173.315 |
| 2 | 116.721 | 136.185 | 155.383 | 175.174 | 194.762 | 233.529 | 262.627 | 350.275 |
| 3 | 175.164 | 204.145 | 232.833 | 262.642 | 291.808 | 350.475 | 394.536 | 496.645 |
| 4 | 233.433 | 272.073 | 310.135 | 349.986 | 388.821 | 466.417 | 525.797 | 642.178 |
| 5 | 291.596 | 339.904 | 387.711 | 437.150 | 485.647 | 583.552 | 656.902 | 778.689 |
| 6 | 349.853 | 407.910 | 465.105 | 524.465 | 582.821 | 699.712 | 787.794 | 905.501 |
| 7 | 407.743 | 475.545 | 542.593 | 611.803 | 679.821 | 815.764 | 919.059 | 1055.601 |
| 8 | 466.308 | 543.538 | 620.122 | 699.267 | 777.270 | 932.921 | 1050.723 | 1206.556 |
| 9 | 524.215 | 611.435 | 697.361 | 786.080 | 874.225 | 1049.313 | 1180.750 | 1356.730 |
| 10 | 582.334 | 679.221 | 775.060 | 873.735 | 970.961 | 1165.853 | 1312.283 | 1507.250 |
| 11 | 640.228 | 747.280 | 852.223 | 961.068 | 1068.297 | 1282.737 | 1442.230 | 1657.677 |
| 12 | 698.779 | 814.622 | 929.632 | 1048.545 | 1165.097 | 1399.411 | 1573.390 | 1809.076 |
| 13 | 757.103 | 883.023 | 1006.790 | 1135.725 | 1262.370 | 1515.016 | 1705.398 | 1959.611 |
| 14 | 814.907 | 950.696 | 1084.856 | 1223.199 | 1359.640 | 1632.194 | 1835.762 | 2110.138 |

*Table 26 L1 cache bandwidth for Intel Xeon E5-2697v3*

*Frequency is given in GHz bandwidth in GB/s*

| | 1.2 | 1.4 | 1.6 | 1.8 | 2 | 2.4 | 2.6 | turbo |
|---|---|---|---|---|---|---|---|---|
| | | | | L2 | | | | |
| 1 | 35.631 | 41.448 | 46.333 | 52.270 | 60.106 | 71.467 | 78.096 | 106.341 |
| 2 | 71.549 | 83.287 | 93.941 | 104.657 | 115.227 | 141.170 | 161.351 | 213.643 |
| 3 | 108.377 | 124.279 | 140.487 | 160.837 | 176.693 | 212.067 | 239.174 | 302.287 |
| 4 | 143.720 | 164.197 | 190.922 | 211.313 | 237.615 | 286.958 | 317.811 | 395.047 |
| 5 | 180.205 | 210.258 | 237.169 | 266.478 | 298.897 | 352.539 | 404.646 | 474.544 |
| 6 | 215.578 | 249.033 | 282.761 | 322.933 | 357.378 | 425.180 | 483.121 | 558.192 |
| 7 | 249.075 | 291.718 | 328.011 | 374.710 | 412.850 | 502.037 | 557.590 | 647.802 |
| 8 | 284.672 | 332.804 | 380.264 | 426.648 | 474.888 | 572.152 | 643.852 | 735.237 |
| 9 | 322.015 | 372.631 | 423.816 | 480.738 | 532.103 | 639.340 | 727.663 | 833.742 |
| 10 | 360.412 | 413.941 | 471.789 | 537.329 | 594.818 | 716.906 | 798.446 | 914.736 |
| 11 | 395.914 | 460.556 | 522.941 | 589.991 | 651.623 | 790.478 | 874.767 | 1021.361 |
| 12 | 429.662 | 500.933 | 566.958 | 644.700 | 708.156 | 849.013 | 957.194 | 1110.313 |
| 13 | 464.580 | 540.836 | 618.899 | 691.894 | 769.998 | 927.418 | 1044.378 | 1204.968 |
| 14 | 499.139 | 582.573 | 664.470 | 745.416 | 830.550 | 996.766 | 1124.870 | 1291.927 |

*Table 27 L2 cache bandwidth for Intel Xeon E5-2697v3*

*Frequency is given in GHz, bandwidth in GB/s*

| | 1.2 | 1.4 | 1.6 | 1.8 | 2 | 2.4 | 2.6 | turbo |
|---|---|---|---|---|---|---|---|---|
| | | | | L3 | | | | |
| 1 | 16.595 | 18.470 | 20.624 | 23.382 | 26.191 | 30.700 | 33.445 | 45.161 |
| 2 | 36.685 | 42.990 | 48.739 | 55.089 | 61.423 | 72.209 | 77.878 | 90.182 |
| 3 | 54.203 | 63.625 | 71.713 | 81.153 | 88.395 | 105.386 | 112.122 | 129.454 |
| 4 | 72.298 | 82.022 | 91.747 | 103.685 | 111.369 | 124.832 | 133.296 | 148.537 |
| 5 | 84.096 | 95.278 | 104.454 | 115.176 | 125.238 | 140.721 | 150.066 | 165.709 |
| 6 | 97.037 | 108.298 | 119.293 | 130.320 | 141.552 | 162.800 | 168.600 | 187.863 |
| 7 | 106.794 | 116.979 | 127.787 | 137.257 | 144.978 | 163.228 | 171.896 | 189.408 |
| 8 | 109.697 | 120.612 | 131.174 | 143.411 | 150.417 | 167.231 | 191.681 | 204.619 |
| 9 | 110.262 | 118.517 | 128.757 | 137.391 | 145.275 | 159.048 | 166.717 | 187.379 |
| 10 | 108.485 | 119.382 | 129.970 | 139.807 | 148.268 | 161.921 | 168.580 | 180.170 |
| 11 | 112.247 | 123.902 | 136.674 | 146.692 | 156.385 | 172.517 | 181.013 | 188.426 |
| 12 | 117.734 | 129.940 | 142.180 | 154.644 | 165.214 | 182.936 | 192.025 | 200.079 |
| 13 | 122.789 | 136.318 | 149.207 | 160.840 | 172.309 | 190.243 | 198.933 | 212.039 |
| 14 | 128.571 | 143.422 | 156.717 | 169.373 | 182.132 | 199.464 | 208.916 | 223.238 |

*Table 28 L3 cache bandwidth for Intel Xeon E5-2697v3*

*Frequency is given in GHz, bandwidth in GB/s*

| | 1.2 | 1.4 | 1.6 | 1.8 | 2 | 2.4 | 2.6 | turbo |
|---|---|---|---|---|---|---|---|---|
| | | | | DRAM | | | | |
| 1 | 8.639 | 8.951 | 9.720 | 10.545 | 11.432 | 12.691 | 13.436 | 16.011 |
| 2 | 18.197 | 19.118 | 19.884 | 20.502 | 22.067 | 24.413 | 29.159 | 30.453 |
| 3 | 27.524 | 29.396 | 30.699 | 31.581 | 33.114 | 34.974 | 39.758 | 40.830 |
| 4 | 38.966 | 41.124 | 43.596 | 45.910 | 47.102 | 49.025 | 50.998 | 50.965 |
| 5 | 49.171 | 51.379 | 52.967 | 54.400 | 55.229 | 56.579 | 57.412 | 57.010 |
| 6 | 54.904 | 56.239 | 57.421 | 57.945 | 58.302 | 58.964 | 59.295 | 58.866 |
| 7 | 57.383 | 57.970 | 58.498 | 58.772 | 58.646 | 59.295 | 59.323 | 59.261 |
| 8 | 57.495 | 57.674 | 58.064 | 57.842 | 57.956 | 58.513 | 58.486 | 58.554 |
| 9 | 56.643 | 57.038 | 57.328 | 57.219 | 57.348 | 57.838 | 57.947 | 57.862 |
| 10 | 56.762 | 56.890 | 57.094 | 57.062 | 57.142 | 57.599 | 57.819 | 57.490 |
| 11 | 56.596 | 56.754 | 57.009 | 56.834 | 57.007 | 57.420 | 57.562 | 57.377 |
| 12 | 56.472 | 56.367 | 56.820 | 56.692 | 57.061 | 57.149 | 57.379 | 57.014 |
| 13 | 56.334 | 56.305 | 56.663 | 56.801 | 56.848 | 57.037 | 57.263 | 56.855 |
| 14 | 56.041 | 56.162 | 56.490 | 56.586 | 56.647 | 56.713 | 56.768 | 56.937 |

*Table 29 DRAM bandwidth Intel Xeon E5-2697v3*

*Frequency is given in GHz, bandwidth in GB/s*

Time required to compute single FLOP – *Tflop* – is given by

$$Tflop = \frac{1}{P(f,n)*U}$$

while time required to send one byte of data – *Tmop* – is given by

$$Tmop = \frac{1}{L1(f,n)*V+L2(f,n)*X+L3(f,n)*Y+DRAM(f,n)*Z}$$

The coefficients for each loop are summarised in Table 30.

| SH TCO639 loop | V | X | Y | Z | U |
|---|---|---|---|---|---|
| ledir->dgemm_327 | 0.0381 | 0.1097 | 0.0201 | 0.0027 | 0.1988 |
| leinv->dgemm_315 | 0.4100 | 5.5113E-05 | 0 | 0.9612 | 0.2683 |
| Trmtol_134 | 2.4472E-05 | 1.8389E-05 | 1.8705E-05 | 0.3637 | - |
| Trltom_130 | 2.4471E-05 | 1.8389E-05 | 1.8705E-05 | 0.3582 | - |
| ftinv_ctl->fourier_in_54 | 2.1605E-07 | 0 | 0.3914 | 0.9997 | - |
| prfi1b_91 | 0.0002 | 0.0033 | 0 | 0.2831 | - |
| ltdir_ctl->updsp_132 | 0.0009 | 0.0013 | 0.0830 | 0.0015 | - |
| asre1b_88 | 0.0009 | 0.0013 | 0.1234 | 0.0015 | 0.0099 |
| Trgtol_434 | 1.5691E-07 | 1.2459E-07 | 0.1014 | 0.6122 | - |
| Trltog_433 | 0 | 9.0057E-07 | 0 | 0.9511 | - |
| ftdir_104 | 0 | 3.6131E-06 | 0.1106 | 1 | - |
| prfi2b_80 | 7.4234E-08 | 5.7067E-08 | 0.0001 | 0.8117 | - |
| fourier_out_53 | 0 | 0 | 0.0360 | 1 | - |
| Leinv_179 | 0.0050 | 0.0149 | 0.0172 | 0.0312 | - |
| Leinv_142 | 0.0010 | 0.0238 | 0.0343 | 0.0015 | - |

*Table 30 Tmop and Tflop coefficients*

The execution time of each loop is calculated using following equation:

$$T_{func} = \max(W * T_{flop}, Q * T_{mop})$$

so that the total execution time of a dwarf is given by

$$T = T_{func1} + \cdots + T_{funcn-1} + T_{funcn}$$

In order to run Spherical Harmonics or other dwarfs in a multinode environment, MPI library is used to exchange data between concurrent parts of a dwarf which run on separate CPU nodes. The computational domain is divided between nodes, so that amount of work *W* and moved data *Q* lowers when adding more nodes – it is divided between MPI processes (i.e. nodes in our case), so that *i-th* node has to process *W/n* work and *Q/n* data, where *n* is the number of nodes.

Next, number of *W* and *Q* per node is calculated for each loop for each node. We assume that each node is equipped with the same processor running at 2.6GHz frequency, and the same type and amount of DRAM memory. Equations to calculate *Tflop* and *Tmop* remain the same, though they allow to capture differences between nodes in case they are equipped with different CPUs. To calculate time required to run part of the SH dwarf on *i-th* node, following equations are used:

$$Tflop_i = \frac{1}{P_i(f,n)*U_i}$$

$$Tmop_i = \frac{1}{L1_i(f,n)*V_i + L2_i(f,n)*X_i + L3_i(f,n)*Y_i + DRAM_i(f,n)*Z_i}$$

$$T_{func_i} = \max(W_i * Tflop_i, Q_i * Tmop_i)$$

$$T_i = T_{func1_i} + \cdots + T_{funcn-1_i} + T_{funcn_i}$$

Table 31 presents *W* and *Q* values for different number of nodes.

| *Metric* | Loop name | Number of nodes | | | |
|---|---|---|---|---|---|
| | | 1 | 2 | 4 | 8 |
| W [FLOP] | ledir->dgemm_327 | 8.70736E+12 | 4.35368E+12 | 2.17684E+12 | 1.08842E+12 |
| | leinv->dgemm_315 | 8.70736E+12 | 4.35368E+12 | 2.17684E+12 | 1.08842E+12 |
| | asre1b_88 | 33228800000 | 16614400000 | 8307200000 | 4153600000 |
| Q [Bytes] | ledir->dgemm_327 | 4.65232E+13 | 2.32616E+13 | 1.16308E+13 | 5.8154E+12 |
| | leinv->dgemm_315 | 1.04509E+14 | 5.22547E+13 | 2.61273E+13 | 1.30637E+13 |
| | Trmtol_134 | 5.31661E+11 | 2.6583E+11 | 1.32915E+11 | 66457600000 |
| | Trltom_130 | 5.31661E+11 | 2.6583E+11 | 1.32915E+11 | 66457600000 |
| | ftinv_ctl->fourier_in_54 | 5.31661E+11 | 2.6583E+11 | 1.32915E+11 | 66457600000 |
| | prfi1b_91 | 5.31661E+11 | 2.6583E+11 | 1.32915E+11 | 66457600000 |
| | ltdir_ctl->updsp_132 | 3.78742E+11 | 1.89371E+11 | 94685465600 | 47342732800 |
| | asre1b_88 | 5.31661E+11 | 2.6583E+11 | 1.32915E+11 | 66457600000 |
| | Trgtol_434 | 7.31034E+11 | 3.65517E+11 | 1.82758E+11 | 91379200000 |
| | Trltog_433 | 6.64576E+11 | 3.32288E+11 | 1.66144E+11 | 83072000000 |

| | | | | |
|---|---|---|---|---|
| ftdir_104 | 5.31661E+11 | 2.6583E+11 | 1.32915E+11 | 66457600000 |
| prfi2b_80 | 5.31661E+11 | 2.6583E+11 | 1.32915E+11 | 66457600000 |
| fourier_out_53 | 5.31661E+11 | 2.6583E+11 | 1.32915E+11 | 66457600000 |
| Leinv_179 | 3.59536E+11 | 1.79768E+11 | 89883904000 | 44941952000 |
| Leinv_142 | 3.59536E+11 | 1.79768E+11 | 89883904000 | 44941952000 |

*Table 31 W and Q values for different number of nodes*

To synchronise work done within each node, message passing approach is used. The amount of data that needs to be exchanged depends on the problem size, implementation of parallelism, type of data and structures moved between nodes, etc. The communication may happen in between loops, in the beginning or end of each iteration, etc, thus we propose to have a cumulative approach to model communication. Execution time at *i-th* node is then:

$$T_i = T_{func1_i} + \cdots + T_{funcn-1_i} + T_{funcn_i} + Tcomm_i$$

The communication part is highly dependent on size of data to be exchanged and communication pattern between parallel parts of a dwarf:
- Inbound/outbound data exchange;
- All-to-all, one-to-all, all-to-one;
- Halo exchange, stencil-type;
- Number of communication per iteration/run.

If communication occurs at each iteration, communication time at *i-th* node is given by:

$$Tcomm_i = Tsingle_i * (Q_{in_i} + Q_{out_i}) * iters$$

where $Tsingle_i$ is time required to send or receive one byte of data, $Qin_i$ is amount of data received, $Qout_i$ is amount of data sent, and iters is the number of dwarf iterations when data exchange occurs.

If all parts of the dwarfs runs on different nodes concurrently, the total execution time on the multinode (*n* nodes) is then the time of the slowest node:

$$T_{multinode} = \max(T_1, T_2, \ldots, T_{n-1}, T_n)$$

To simulate execution of the Spherical Harmonics dwarf on multinode architecture, and to generate time-to-solution graphs, DCworms is used. The simulator provides capabilities of assessing the computational and energy performance of NWP dwarfs on different hardware architectures, and the aforementioned multinode performance model has been applied to.

We run test using TCO639 testcase: 16km resolution, 100 iterations, 200 fields, on multinode equipped with Intel E5-2697v3 processors. Please note that we assume a perfect overlap between computation and internode communication

Figure 14 presents comparison between the profiling data and the multinode performance model, which was applied to the DCworms simulator.

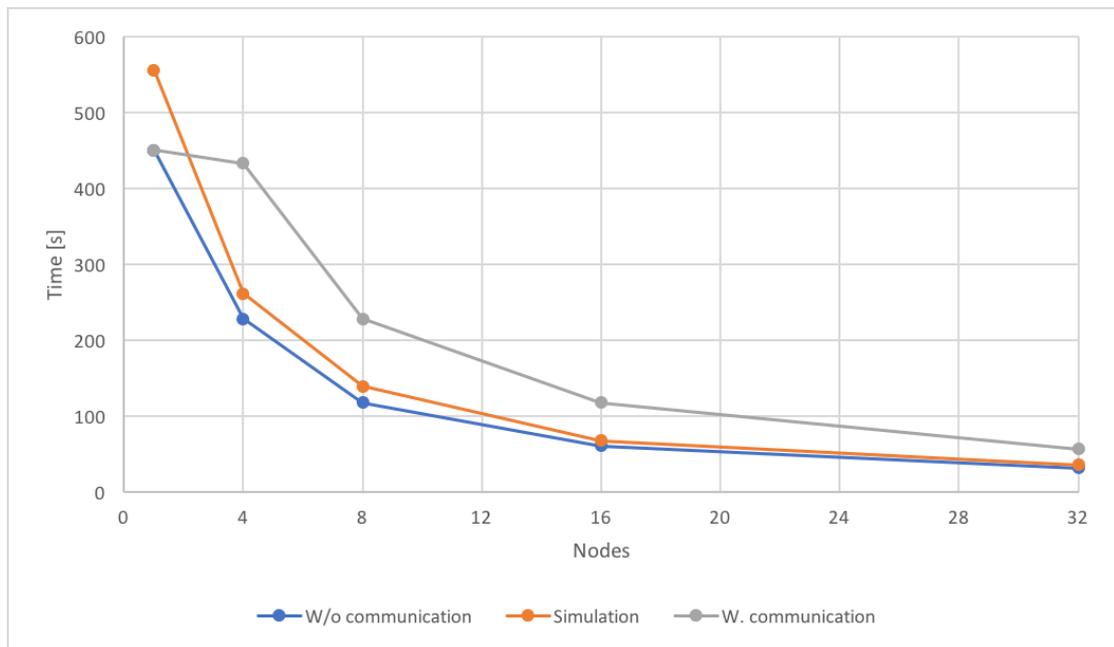

*Figure 14 SH TCO639 time-to-solution: simulation vs. real run*

To validate the model different quality metrics are computed (numbers in brackets when communication time is added):
- The maximum difference is 23% (62%);
- The minimum difference is 11% (23%);
- The mean difference is 15.6% (52.4%);
- The standard deviation is 4.92% (16,5%).

## 5 Energy modelling

### 5.1 State-of-the-art

The energy model allows to analyse, model and predict application energy based on a processing unit's computation and communication capability. In (Choi, Bedard, Fowler, i Vuduc, 2013), authors have assumed that each operation has a fixed energy cost and a fixed data movement cost, while the constant energy cost is linear in time. The constant power usage depends on both, hardware and algorithm, and includes both static and leakage power management. However, the proposed model does not include dynamic power management of charging and discharging gate capacitance. The authors assumed that time per work (arithmetic) operation and time per memory operation are estimated with the hardware peak throughput values, whereas the energy cost is estimated using a linear regression based on real experiments.

Another set of extensions to the Roofline model have been proposed in (Hager, Teribig, Habich, i Wellein, 2014) to model energy on dual multi-core CPU with three levels of cache hierarchy. In this approach, the dynamic power management was modelled as a second degree polynomial, based on real benchmark data, that scales linearly with the number of active cores up to the saturation point. The authors assumed that dynamic power depends quadratically on the frequency. In the saturation point, energy to solution ratio grows with the number of used cores, that is proportional to dynamic power, while the time to solution remains constant. This approach is described in details in Deliverable D4.5 (Section 3.2).

### 5.2 Model of energy

The energy is modelled similarly to modelling the performance in the roofline model. The model of energy is based on the notion of useful work $W$ and the number of moved bytes $Q$ through the slowest data path. As for the roofline model, the work is represented by the number of floating-point operations being executed $W$. This model assumes that calculation of every FLOP and sending the single byte of memory cost some energy. Moreover, the processing unit during execution of the code draws some power. The following equation is used to calculate the energy cost of application execution on the processing unit:

$$E = W * e_{flop} + Q * e_{mop} + T * P0$$

where $e_{flop}$ is the energy cost per FLOP, $e_{mop}$ is energy cost per single byte moved (MOP - memory operation) and $P0$ is the constant power draw. The constant power is based on P-state of the processing unit and the number of cores used. For the energy usage measurements some hardware vendors expose API. For example, the Running Average Power Limit (RAPL) interface is exposed for Intel CPUs, whereas Nvidia shares the NVML (Nvidia Management Library) interface.

RAPL provides a set of counters giving energy and power consumption information. It estimates energy usage by using hardware performance counters. Energy estimation is done for up to four domains, see Figure 15 :
- PKG – entire package;
- Cores;
- Uncore, e.g. GPU, caches (not available on all processors);

- DRAM – main memory (not available on all processors).

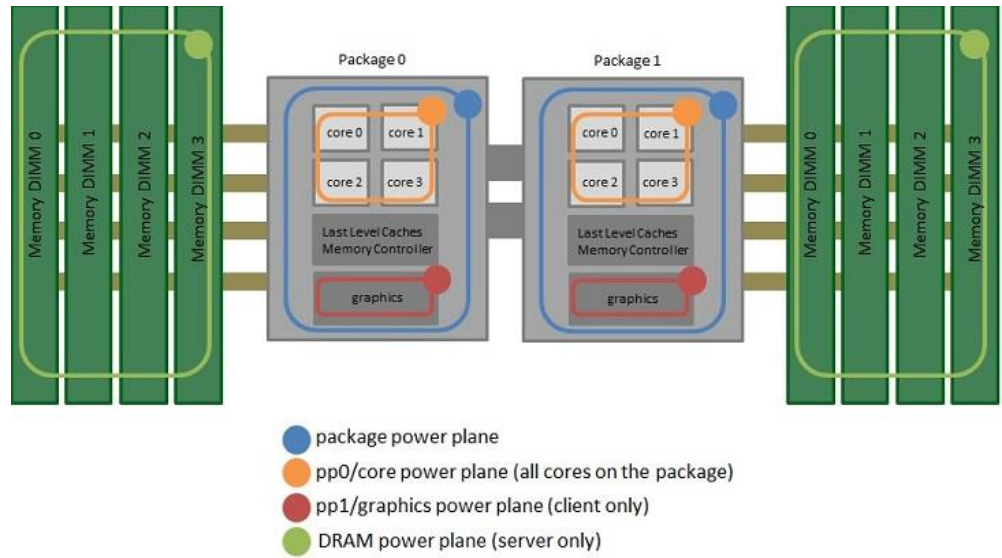

*Figure 15 Available power domains*

Our energy model uses only PKG and DRAM information, because i) Uncore information is not available for every processor, and ii) the following relationships holds: Cores + Uncore <= PKG.

To measure energy consumption, we use rapl_read tool (http://web.eece.maine.edu/~vweaver/projects/rapl/index.html), that allows to measure energy using:
- powercap interface (introduced in Linux 3.13);
- perf_event interface (Linux 3.14 and newer);
- raw access to MSRs (Model Specific Registers).

Rapl_read tools provides such information:

```
$ ./rapl-read -s

RAPL read -- use -s for sysfs, -p for perf_event, -m for msr

Found Skylake Processor type
        0 (0), 1 (0), 2 (0), 3 (0),
        Detected 4 cores in 1 packages

Trying sysfs powercap interface to gather results

        Sleeping 1 second

        Package 0
                package-0       : 1.606746J
                core    : 0.327941J
                uncore  : 0.040039J
                dram    : 0.481933J
```

The code has been modified to allow for finer time resolution, e.g. 100ms, and to measure energy of a given application.

Similar to the roofline performance model, we use benchmarks to provide energy consumption table for given architecture. To measure DRAM energy consumption, we use lmbench tool (http://lmbench.sourceforge.net/). This tool has to be compiled with correct flags to obtain the correct measurements:
```
cd lmbench-3.0-a9/src
```

All occurrences of CFLAGS have to be changed in Makefile to:
```
CFLAGS="-O -xHost"
```

Next, lm_bech is compiled:
```
Make
```

To run the benchmark the following commands are executed:
```
cd bin/x86_64-linux-gnu
./bw_mem –P $[t} –N 5 –W 5 1000m fwr
```

where –P is the number of threads used, -N is the number of repetitions, -W is the number of warm-up runs and 1000m is the amount of memory to be moved through memory (m suffix stands for MB). A *fwr* test is used (full write) because it requires more energy than full read (*frd*) test. It is important to test memory larger that L3 cache size, otherwise bandwidth of caches will be tested, and energy consumption will be captured for PKG/Uncore domain instead of DRAM.

For measuring PKG (including cores) energy, we use aforementioned rapl_read tool with a CPU stress test application *mprime* (https://www.mersenne.org/). *Mprime* is a popular choice for stress testing a CPU, and allows for generating maximum power consumption (using in-place large FFTs test). Table 32 presents measured PKG and DRAM energy per number of utilised cores for Intel i7-4700EQ.

| CORES | ENERGY [J] | |
|---|---|---|
| | PKG | DRAM |
| IDLE STATE | 5.53 | 1.37 |
| 1 | 23.28 | 4.73 |
| 2 | 35.12 | 5.42 |
| 3 | 42.85 | 5.89 |
| 4 | 49.71 | 7.21 |

*Table 32 Energy usage for Intel i7-4700EQ*

RAPL is not a power meter, it rather uses a software power model, providing estimated energy measurements. Because of its nature, we propose a slightly modified energy model. It still relies on $W$ - number of calculated FLOP, and $Q$ – number of bytes moved through memory hierarchy. Instead of trying to find $e_{flop}$ - energy of one FLOP, and $e_{mop}$ – energy of one byte moved, which depends greatly on e.g. CPU architecture, memory DRAM hardware specification, operation type, type of access to memory, we estimate energy based on time required to perform work $W$ and send $Q$ bytes of data.

Likewise in the performance model, we assume the communication perfect overlaps computation. The energy model is then following:

$$Esingle = Epkg(f,n) * W * T_{flop} + Edram(f,n) * Q * T_{mop}$$

Constant power draw of PKG and DRAM has to be included, so that the energy model is:

$$Esingle = Epkg(f,n) * W * T_{flop} + Edram(f,n) * Q * T_{mop} + Econst$$

where
$Epkg(f,n)$ is measured PKG energy for given CPU frequency and number of threads,
$W$ is number of FLOP to be calculated,
$T_{flop}$ is time required to calculate one FLOP,
$Edram(f,n)$ is measured memory energy for given DRAM, CPU frequency and number of threads,
$Q$ is the amount of bytes to be sent,
$T_{mop}$ is the time required to send one byte of data through memory hierarchy,
and

$$Econst = \begin{cases} Edram(f,idle) * (W * Tflop - Q * Tmop); W * Tflop > Q * Tmop \\ Epkg(f,idle) * (Q * Tmop - W * Tflop); W * Tflop < Q * Tmop \\ 0; W * Tflop = Q * Tmop \end{cases}$$

### 5.3 Correlation of the processing unit

The last step is to correlate parameters of the processing unit with the dwarf. For the PKG part this is done by finding coefficients U and S by using linear regression:

$$Epkg(f,n) * U + S * Epkg(f,idle) = Empkg(f,n)$$

where *Epkg* is the energy of the entire package under full load and *Empkg* is the measured energy consumption.

For the DRAM part the coefficients X and Y are found by using linear regression:

$$Edram(f,n) * X + Y * Edram(f,idle) = Emdram(f,n)$$

where *Edram* is the energy of DRAM under full write and *Emdram* is the measured DRAM energy. Table 33 presents example PGK energy coefficients for Spherical Harmonics dwarf, while Table 34 presents coefficients for DRAM energy.

| CORES | EPKG [J] | EMPKG [J] | U | S |
|---|---|---|---|---|
| IDLE STATE | 5.53 | | | |
| 1 | 23.28 | 18.6 | | |
| 2 | 35.12 | 25.04 | 0.58309038 | 0.50242954 |
| 3 | 42.85 | 28.31 | | |
| 4 | 49.71 | 32.26 | | |

*Table 33 PKG energy coefficients for SH dwarf*

| CORES | EDRAM [J] | EMDRAM [J] | X | Y |
|---|---|---|---|---|
| IDLE STATE | 1.31 | | | |
| 1 | 4.73 | 1.77 | | |
| 2 | 5.42 | 2.3 | 0.37420719 | 0.5 |
| 3 | 5.89 | 2.8 | | |
| 4 | 7.21 | 3.4 | | |

*Table 34 DRAM energy coefficients for SH dwarf*

### 5.4 CPU validation

To validate the model for CPU architecture, Spherical Harmonics dwarf has been tested on another CPU architecture, Intel E5-2640v3 with 8 physical cores. The comparison between measured and modelled PKG + DRAM energy is presented in Table 35.

| cores | Epkg [J] | Edram [J] | Time [s] | Epkg model [J] | Edram model [J] | Epkg measured [J] | Edram measured [J] | Total energy model [J] | Total energy measured [J] |
|---|---|---|---|---|---|---|---|---|---|
| idle | 31.82 | 3.71 | | | | | | | |
| 1 | 44.89 | 11.1 | 139.9 | 5898.49 | 840.61 | 5614 | 1085 | 6739.11 | 6699 |
| 2 | 54.23 | 14.51 | 71.2 | 3389.71 | 518.67 | 3124 | 624.8 | 3908.38 | 3748.8 |
| 4 | 73.93 | 18.36 | 36.9 | 2180.61 | 321.96 | 1955.7 | 354.24 | 2502.58 | 2309.94 |
| 8 | 101.23 | 19.89 | 20.1 | 1507.77 | 186.88 | 1340 | 236 | 1694.66 | 1576 |

*Table 35 Measured vs. Modelled energy for SH dwarf*

- The maximum difference is 7.7%;
- The minimum difference is 0.6%;
- The mean difference is 7.7%;
- The standard deviation is 3.24%.

### 5.5 GPU consideration

In order to provide a very detailed energy model for GPUs, various components need to be considered, e.g. hardware utilisation, rail voltages, currents leakage [Stokke]. We proposed a simplified approach to model energy consumption on GPU accelerators. The GPU energy consumption may read as a Uncore domain field from the RAPL interface. However, this information is not always available. For measuring GPU energy we use Nvidia's System Management Interface: nvidia-smi. Depending on the card generation, various levels of information can be gathered. To monitor current power draw of GPU card it is sufficient to issue following command:

```
nvidia-smi stats -i $CUDA_VISIBLE_DEVICE  -d pwrDraw
```

where $CUDA_VISIBLE_DEVICE is index of GPU card to monitor.

For the GPU energy model, we take into consideration energy accounted to package to which GPU is connected:

$$E = T * (Epkg(f, idle) + S * Egpu)$$

where *E* is the total energy, T is the time required to perform calculations, *Epkg(f,idle)* is the constant CPU power draw at idle state, *Egpu* is the GPU power limit and *S* if the fraction of GPU power limit used by considered algorithm of application.

Table 36 provides details on GPU energy usage for various accelerators. The measurements were taken running BiFFT at different grid sizes. Because power consumption fluctuates during dwarf execution, we provide average energy consumption per second. Power rating ranged from 55W to 104W and from 47W to 76W for GeForce 970 and Tesla K20 respectively. We expect that power draw changes during execution because of certain activity and inactivity time for which GPUs SM waits for data to be available and due to synchronisations. Similar issue is observed in multi-GPU environment, as described in Deliverable D4.5 (Section 3).

| GPU | Tesla K20 | GeForce GTX970 |
|---|---|---|
| **Power limit [W]** | 225 | 155 |
| **AVG energy [W/s]** | 65 | 91 |
| **S [% of power limit]** | 28.89% | 58.71% |

*Table 36 ACRANEB2 energy consumption*

### 5.6 Multinode consideration

Energy modelling on multinode machines is similar to modelling on single CPU or a node. We assume that energy accounted to transferring data between nodes is already estimated in PKG RAPL reading. Thus, energy model for a single dwarf on multinode system is modelled with the following equation:

$$Emulti = \sum_{i=1}^{n} Esingle$$

where *Esingle* is the energy consumption on a single node, *n* is the number of nodes.

### 5.7 NWP consideration

NWP applications run over number of nodes and cores, including accelerators e.g. GPUs. Such applications are modelled using different dwarfs working together, so that a workflow of interaction (e.g. communication) can be described. Moreover, each dwarf may be run on a different hardware architecture. To this end, we propose two equations to calculate the NWP application energy, depending on possible overlap between different dwarfs used. Please note that these equations are based on two dwarfs working together, more advanced examples will require slight modifications based on the nature of the application.

If there is no overlap possible, e.g. one dwarf (running on architecture 1) is waiting for the other (running on architecture 2) to complete its calculations, the energy equation may be following:

$$Enwp_{nooverlap} = E1 + E1const * T2 + E2 + E2const * T1$$

where *E1* is the energy of dwarf 1, *E2* is the energy of dwarf 2, *E1const* is the constant power draw of architecture 1, *E2const* is the constant power draw of architecture 2, *T1* is the total execution time of dwarf 1, and *T2* is the total execution time of dwarf 2.

In a situation when an overlap may occur, e.g. 2 dwarfs are running simultaneously, the total energy is calculated using following equation:

$$Enwp_{overlap} = E1 + E2 + Econst$$

$$Econst = \begin{cases} (T2 - T1) * E1const; T1 < T2 \\ (T1 - T2) * E2const; T1 > T2 \\ 0; T1 = T2 \end{cases}$$

where *E1* is the energy of dwarf 1, *E2* is the energy of dwarf 2, *T1* is the total execution time of dwarf 1, *T2* is the total execution time of dwarf 2, *E1const* is the constant power draw of architecture 1, *E2const* is the constant power draw of architecture 2.

# 6 Performance at scale

This section outlines how DCworms simulator and aforementioned performance and energy models may be used to assess performance at scale. The simulator allows for estimating computational performance and energy efficiency of dwarfs on current and future hardware architectures. To project the performance, an estimation of memory bandwidth for memory hierarchy is needed, as well as estimation of computation unit performance. For existing architectures, these values may be depicted using available benchmarks tools. For architectures yet to come, they are calculated using hardware specification details. Please refer to Deliverable 3.2 for further details.

To project energy efficiency on current architectures, one has to profile energy characteristic of the processing unit, as described in Section 4.3 of this Deliverable. For architectures yet to come, the energy characteristic has to be estimated or derived from technical specification of the processing unit (e.g. GPU power limit).

We propose time-to-solution and energy-to-solution metrics as a basis to project the performance and energy efficiency at system scale. Further metrics can be formulated, e.g. best computational efficiency, minimal energy usage, trade-off between performance and energy consumption.

## 6.1 Performance and energy efficiency projection of dwarfs

### 6.1.1 Spherical Harmonics

DCworms simulator is used to predict computational performance and energy efficiency of Spherical Harmonics dwarf. We model TL159 testcase running on Intel Xeon E5-2697v3 to depict time-to-solution and energy-to-solution. The number of iterations was equal to 100, while the number of fields was 274. Figure 16 presents the results. Time-to-solution decreases with number of nodes used, but energy-to-solution remains the same. This is possible because our model makes an assumption, that internode and intranode communication is perfectly overlapped by computations. However, the performance of Spherical Harmonics dwarf in multinode environment is very much driven by communication, as stated in Deliverable D3.3 and D3.4. It will be also presented later in this Deliverable.

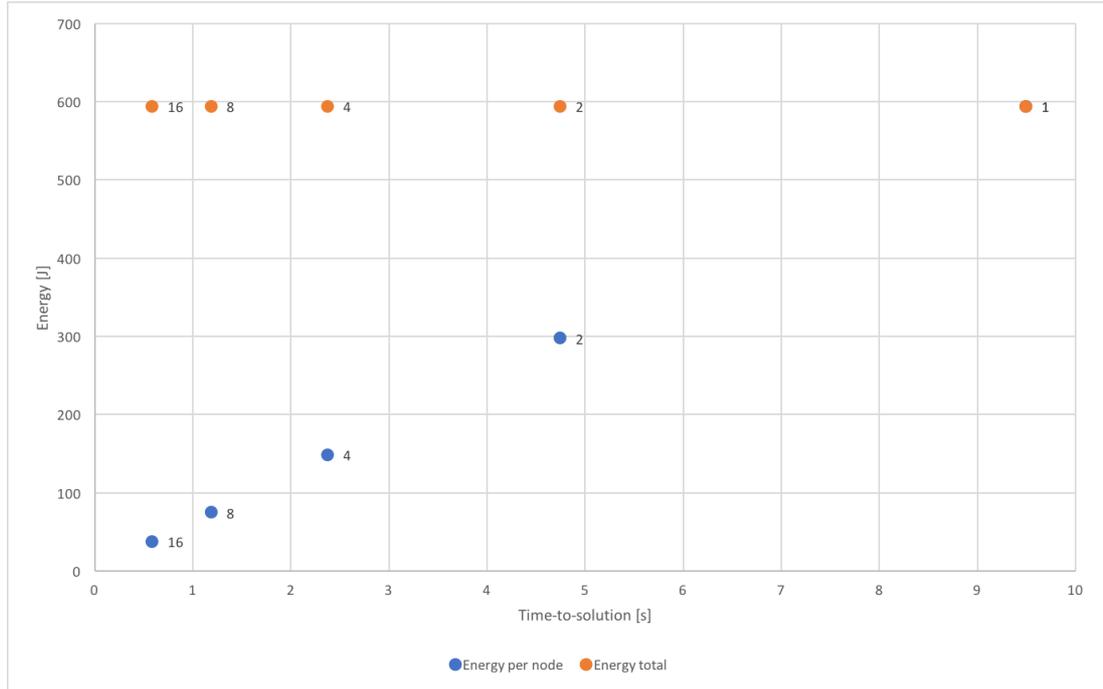

*Figure 16 SH TL159 time-to-solution and energy-to-solution on Intel Xeon E5-2697v3*

*Indices denote number of nodes.*

The Skylake microarchitecture was launched in August 2015 to succeed the Broadwell. It supports AVX-512 instructions known from Intel Knight Landing architecture. In order predict performance of dwarfs running on Skylake CPUs, bandwidth and performance of the processing unit has to be depicted, see Table 37.

| CORES | L1 [GB/S] | L2 [GB/S] | L3 [GB/S] | DRAM [GB/S] | PERFORMANCE [GFLOP/S] |
|---|---|---|---|---|---|
| 1 | 123.78 | 77.19 | 33.41 | 26.61 | 76.8 |
| 2 | 24.,2 | 153.93 | 65.94 | 53.81 | 153.6 |
| 4 | 494.43 | 307.77 | 131.26 | 84.25 | 307.2 |
| 8 | 988.66 | 615.82 | 262.63 | 150.33 | 614.4 |
| 16 | 1976.89 | 1231.29 | 512.84 | 261.01 | 1228.8 |
| 20 | 2470.72 | 1538.71 | 633.32 | 292.84 | 1536 |
| 30 | 3704.44 | 2305.55 | 800.25 | 330.33 | 2304 |
| 40 | 4937.12 | 3073.88 | 836.81 | 309.77 | 3072 |

*Table 37 Intel Xeon Gold 6148 2.4GHz performance and memory bandwidth*

The performance model for Spherical Harmonics dwarf has been already proposed in Deliverable D3.2. Here we provide cache-aware roofline graphs for Skylake for comparison to older microarchitectures. We performed projection for 2 testcases: TL159 (125km resolution, 100 iterations, 274 fields) and TCO639 (16km resolution, 100 iterations, 200 fields).

The roofline comparison between Skylake (Figure 17) and Haswell (Figure 18) microarchitectures tells us that we should expect better time-to-solution for this new microarchitecture.

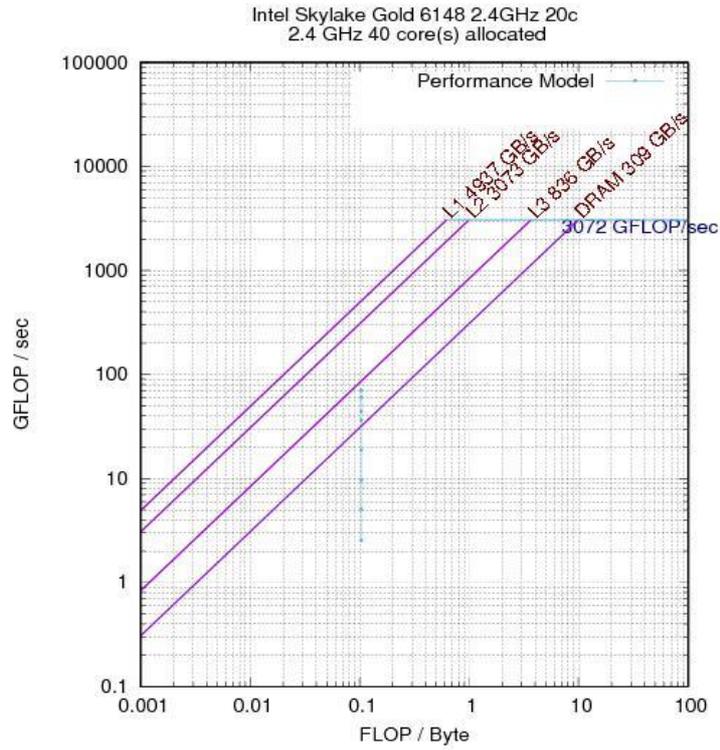

*Figure 17 Cache-aware roofline for Intel Skylake 6148*

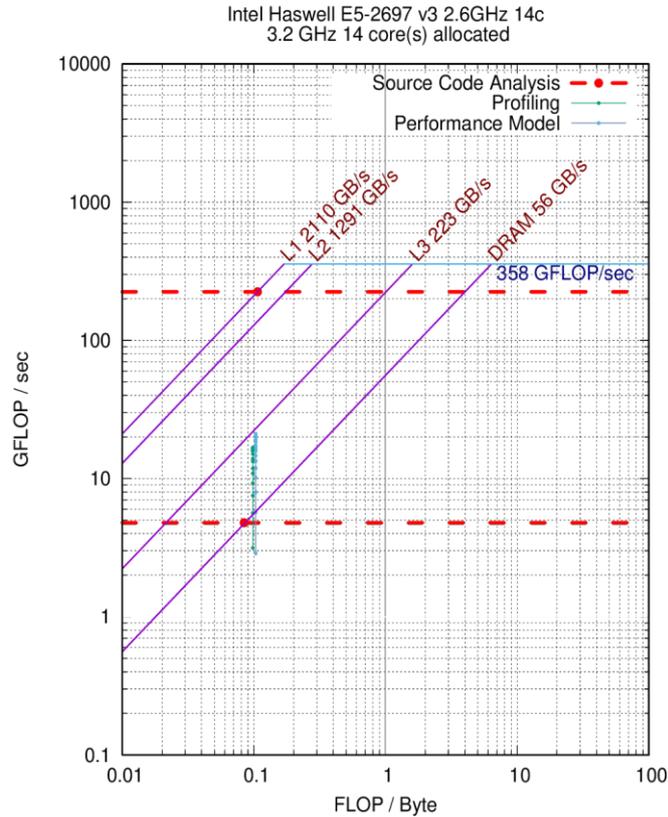

*Figure 18 Cache-aware roofline for Intel E5-2697v3*

DCworms simulation results show that, indeed, the new Skylake microarchitecture performs better with respect to time-to-solution metric, see Figure 19. However, please have in mind that such simulation does not take into account internode communication, thus real differences between runs on different architectures may differ.

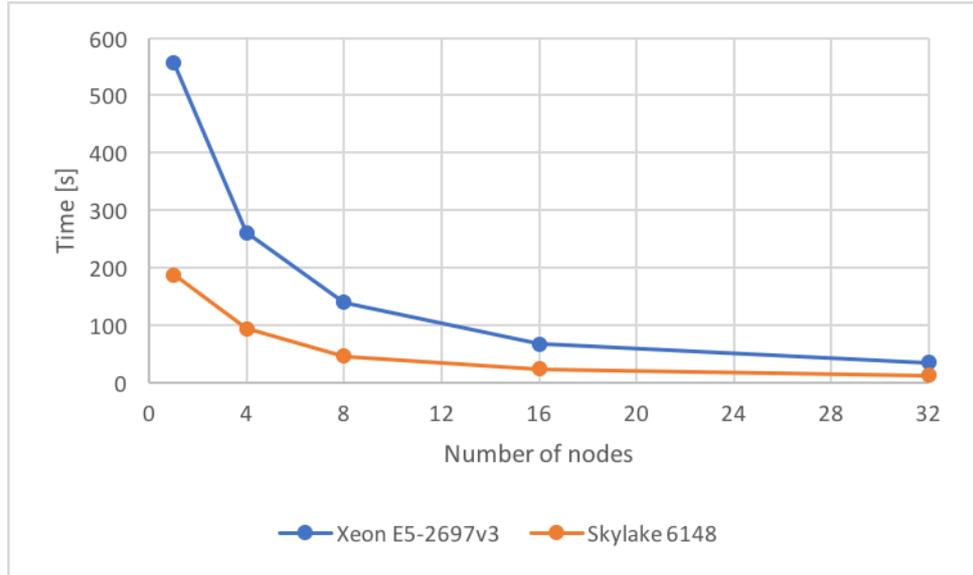

*Figure 19 Spherical Harmonics dwarf, TCO639 testcase, Xeon vs. Skylake simulation*

### 6.1.2 BiFFT

The Bi-Fourier dwarf provides the LAM equivalent of Spherical Harmonics dwarf. It is an essential building block of spectral LAMs such as the ALADIN system. The dwarf implements the spectral transform on a rectangular domain. The spectral transforms are performed consecutively in the zonal and meridional directions. In each of these directions, the transforms are Fast Fourier Transforms, equivalent to the zonal transforms of Spherical Harmonics.
BiFFT dwarf provides the same distribution as the ALADIN LAM:
- In spectral space, the distribution is along the wavenumber and different fields;
- In gridpoint space, the distribution is along the zonal and meridional directions.

Example gridpoint distribution is presented on Figure 20. Further details of the Bi-Fourier spectral transform and the implementation of the dwarf can be found in Deliverable D1.1 (Section 4.4).

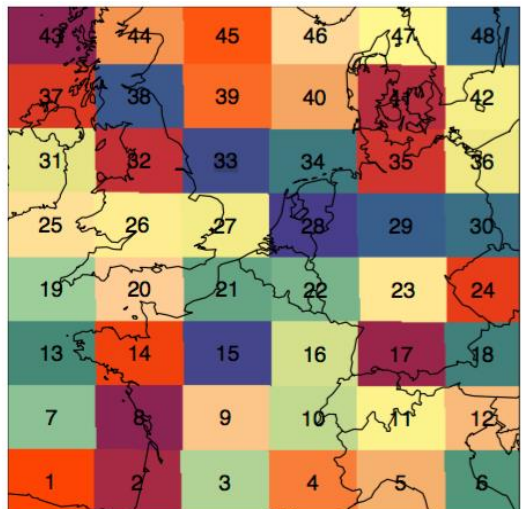

*Figure 20 Gridpoint distribution of a rectangular LAM domain over 48 MPI tasks*

#### 6.1.2.1 Profiling

For the BiFFT dwarf, one direct and one inverse transformation are performed per loop iteration for the prescribed number of fields. In the ALARO reference configuration, the inverse transformations are performed on the number of fields compared to the number of direct transformations; 350 direct ones versus 700 inverse ones. Therefore, to level the playing field we chose to set-up the BiFFT dwarf to perform calculations for 525 fields, resulting in the same total number of spectral transformations performed compared to the ALARO ref. conf. We performed various tests, including input domain size of 200x180 and 800x720, using 20 and 525 fields. These tests were executed on a cluster machine, equipped with Intel Xeon E5-2697v3 processor (see Table 21 for details). Dwarf is run with the OpenMP parallelisation model and with Hyper-Threading technology switched off, running at nominal CPU frequency.

| TESTASE | GFLOP | GB | TIME [S] | I | GFLOP/S | GB/S |
|---|---|---|---|---|---|---|
| 200X180_20 | 4.33 | 32.81 | 1.76 | 0.13 | 2.46 | 18.65 |
| 200X180_525 | 114.30 | 865.04 | 117.72 | 0.13 | 0.97 | 7.35 |
| 800X720_20 | 97.53 | 786.16 | 72.10 | 0.12 | 1.35 | 10.90 |

Table 38 BiFFT execution on a single core

| TESTASE | GFLOP | GB | TIME [S] | I | GFLOP/S | GB/S |
|---|---|---|---|---|---|---|
| 200X180_20 | 4.33 | 32.81 | 1.86 | 0.13 | 2.33 | 17.64 |
| 200X180_525 | 107.21 | 834.80 | 132.96 | 0.13 | 0.81 | 6.28 |
| 800X720_20 | 97.54 | 786.27 | 77.98 | 0.12 | 1.25 | 10.08 |

Table 39 BiFFT execution on multi-core

Table 38 and Table 39 present results for different testcases. All testcases have the computational intensity of 0.12-0.13, disregarding number of used cores, thus we can expect that the code is bounded by memory and not by computations. Another problem lies in poor utilisation of OpenMP parallelism, which requires further investigation and optimisation. This problem has been discussed in Deliverable D3.3 (Section 7). The rooflines for single and multi-core run are presented on Figure 21 and Figure 22 respectively.

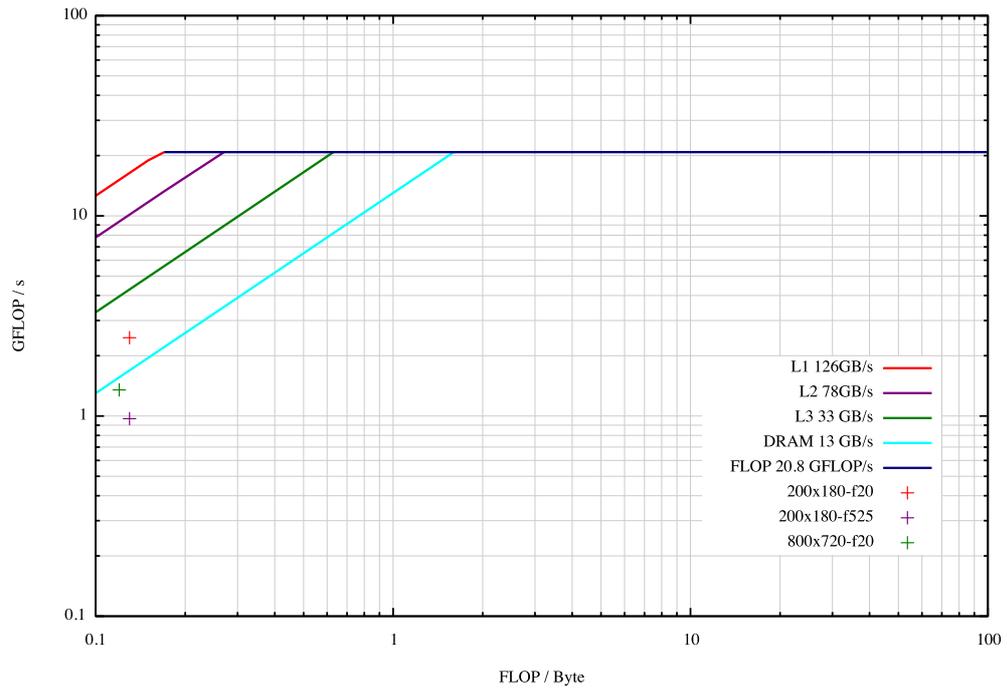

*Figure 21 BiFFT roofline on single core*

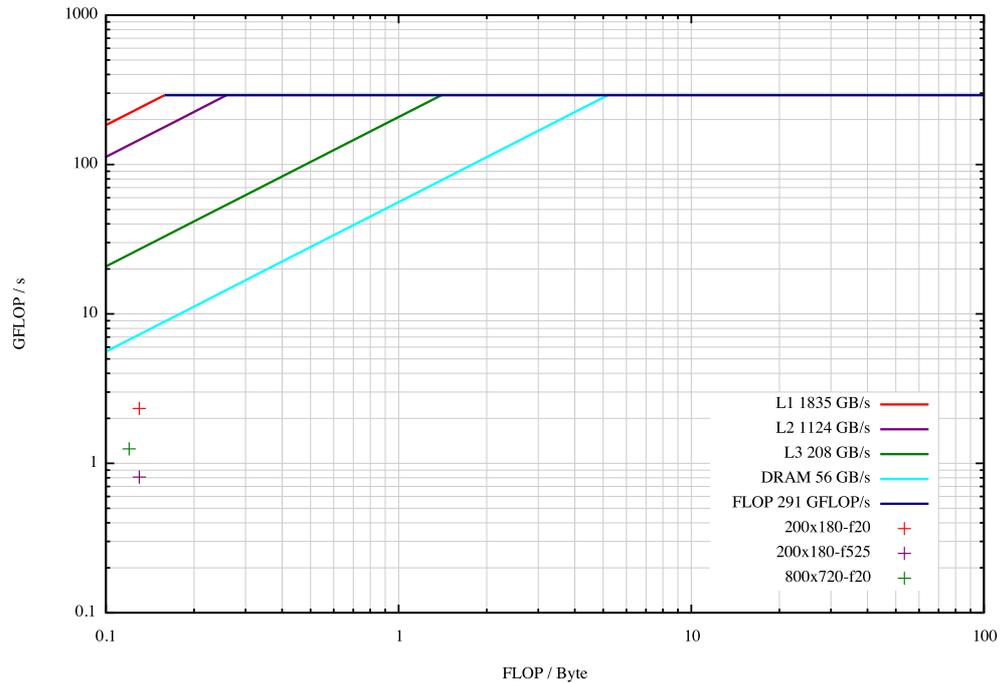

*Figure 22 BiFFT roofline on multi-core*

#### 6.1.2.1 Model instantiation

We have used Intel Advisor tool to profile BiFFT in details. The goal of this action is to deliver a CPU model for DCworms simulator, as described in Deliverable D3.2. Next, most time-consuming loops has been selected for further analysis, see Figure 23.

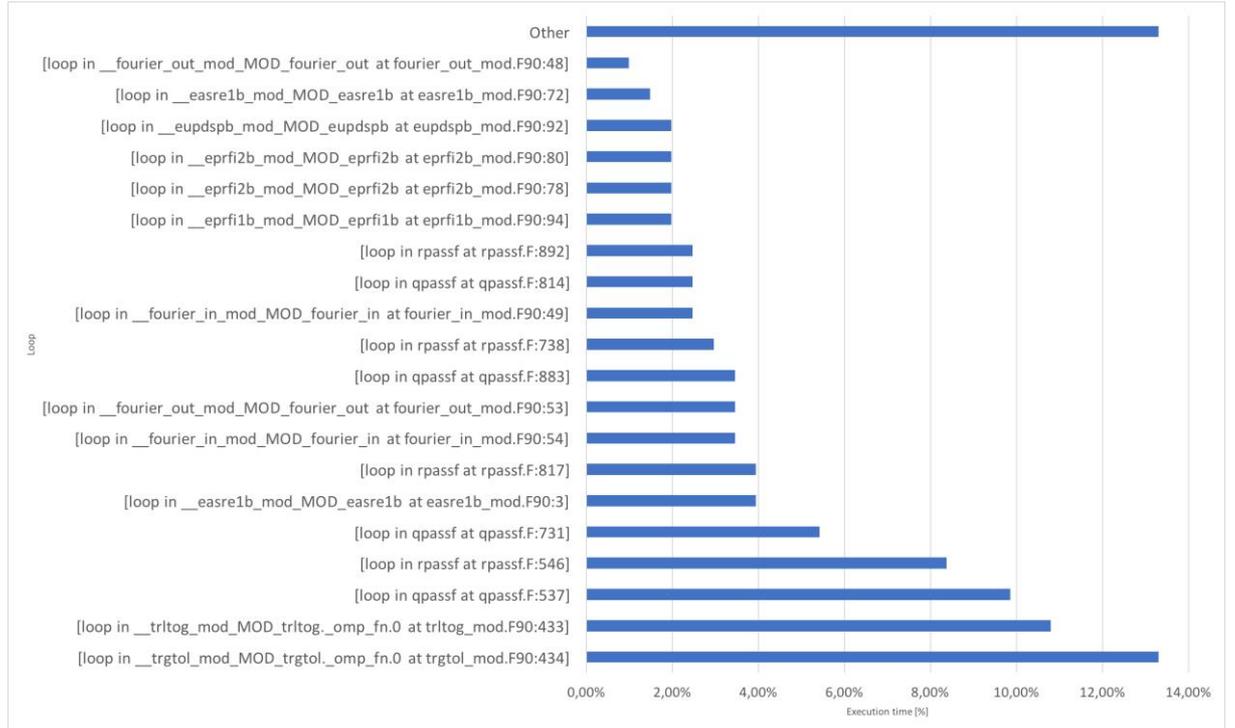

*Figure 23 BiFFT most time-consuming loops*

As seen in 6.1.2.1, computational intensity does not change for different testcases, number of fields, nor number of threads. To correlate the processing unit with a dwarf, different coefficients are to be found as described in Deliverable D3.2. We remind the equations here for convenience. For the memory part this is done by finding coefficients *V*, *X*, *Y* and *Z* that solves the equation:

$$L1(f,n) * V + L2(f,n) * X + L3(f,n) * Y + DRAM(f,n) * Z = R(f,n)$$

where *f* is the frequency, *n* is the number of cores and *R* is the measured bandwidth of a loop based on the state *(f,n)* of the processing unit. The number of equations is equal to the number of tested frequencies multiplied by number of cores. The linear programming (LP) method is used to find the coefficients. For the compute part the *U* and *S* coefficients are designated by using the linear regression:

$$P(f,n) * U + S = M(f,n)$$

where *P* is the performance of a processor and *M* is the performance achieved by loop based on the state *(f, n)*.

In order to calculate execution time for each loop following equation is used:

$$T_{func} = \max(W * T_{flop}, Q * T_{mop})$$

To calculate the execution time of a dwarf the sum of execution time of each function is used:

$$T = T_{func1} + \cdots + T_{funcn-1} + T_{funcn}$$

Taking into account aforementioned equations, V, X, Y, Z, U and S coefficients are depicted. There are 8 loops that employ some computations, whereas other 12 just move data between different structures. Table 40 shows coefficients to calculate time needed to send a single byte of data – Tmop. To calculate time needed to compute single FLOP – Tflop – coefficients given in Table 41 are used.

| loop | V | X | Y | Z | Tmop |
|---|---|---|---|---|---|
| [loop in __easre1b_mod_MOD_easre1b at easre1b_mod.F90:3] | 5.40858E-06 | 0 | 0.001474617 | 0 | 1/(L1(f,n) * V + L2(f,n) X + L3(f,n)*Y + DRAM(f,n)*Z) |
| [loop in __easre1b_mod_MOD_easre1b at easre1b_mod.F90:72] | 0.069381464 | 0 | 0 | 0.067998614 | |
| [loop in __eprfi1b_mod_MOD_eprfi1b at eprfi1b_mod.F90:94] | 0.03717369 | 0 | 0 | 0.133930245 | |
| [loop in __eprfi2b_mod_MOD_eprfi2b at eprfi2b_mod.F90:78] | 0 | 0 | 0 | 0.160125423 | |
| [loop in __eprfi2b_mod_MOD_eprfi2b at eprfi2b_mod.F90:80] | 0.005771743 | 0 | 0.153231217 | 0 | |
| [loop in __eupdspb_mod_MOD_eupdspb at eupdspb_mod.F90:92] | 0.002554603 | 0 | 0 | 0.07466103 | |
| [loop in __fourier_in_mod_MOD_fourier_in at fourier_in_mod.F90:49] | 0.000573908 | 0 | 0 | 0.013448446 | |
| [loop in __fourier_in_mod_MOD_fourier_in at fourier_in_mod.F90:54] | 0.033571136 | 0 | 0 | 0.233430528 | |
| [loop in __fourier_out_mod_MOD_fourier_out at fourier_out_mod.F90:48] | 0 | 0 | 0 | 0.190721296 | |
| [loop in __fourier_out_mod_MOD_fourier_out at fourier_out_mod.F90:53] | 0 | 0 | 0.196896377 | 0.070765159 | |
| [loop in __trgtol_mod_MOD_trgtol._omp_fn.0 at trgtol_mod.F90:434] | -0.004845871 | 0 | 0 | 0.034015643 | |
| [loop in __trltog_mod_MOD_trltog._omp_fn.0 at trltog_mod.F90:433] | -0.009930155 | 0 | 0 | 0.391865155 | |
| [loop in qpassf at qpassf.F:537] | 0.068952513 | | 0.428275464 | 0 | |
| [loop in qpassf at qpassf.F:731] | 0 | 0 | 0.283667875 | 0.824183382 | |
| [loop in qpassf at qpassf.F:814] | 0.020106483 | 0 | 0.037906539 | 0 | |
| [loop in qpassf at qpassf.F:883] | 0.107303836 | 0 | 0.275401098 | 0 | |
| [loop in rpassf at rpassf.F:546] | -0.036361761 | 0 | 0 | 1.434743838 | |
| [loop in rpassf at rpassf.F:738] | 0.038584294 | 0 | 0 | 0.458452275 | |
| [loop in rpassf at rpassf.F:817] | 0.017965398 | 0 | 0 | 0.167872845 | |
| [loop in rpassf at rpassf.F:892] | 0.063895443 | 0 | 0 | 0.737407947 | |

*Table 40 BiFFT Tmop coefficients*

| loop | U | S | Tflop |
|---|---|---|---|
| [loop in qpassf at qpassf.F:537] | 0.116761945 | 1.334821538 | 1/(P(f,n) * U+S) |
| [loop in qpassf at qpassf.F:731] | 0.049320747 | 2.181598462 | |
| [loop in qpassf at qpassf.F:814] | 0.030062493 | 0.245802154 | |
| [loop in qpassf at qpassf.F:883] | 0.20090895 | 1.148413846 | |

| | | |
|---|---|---|
| [loop in rpassf at rpassf.F:546] | 0.056856065 | 1.643513846 |
| [loop in rpassf at rpassf.F:738] | 0.054779697 | 0.762452308 |
| [loop in rpassf at rpassf.F:817] | 0.029270762 | 0.332694154 |
| [loop in rpassf at rpassf.F:892] | 0.122388351 | 1.663112308 |

*Table 41 BiFFT Tflop coefficients*

Next, DCworms is used to create the cache-aware roofline plot. Figure 24 presents comparison between the roofline and cache-aware roofline model for BiFFT running on a single core (200x180-525 fields use case). Figure 25 presents similar comparison for multi-core environment.

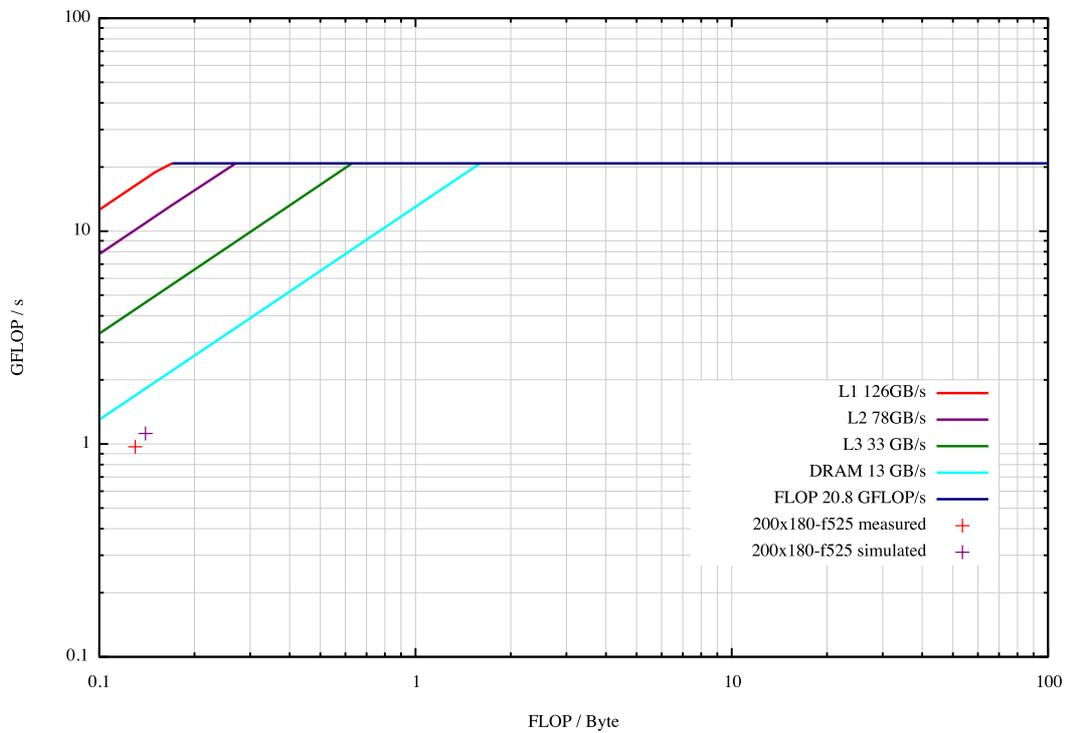

*Figure 24 BiFFT cache-aware roofline on single core*

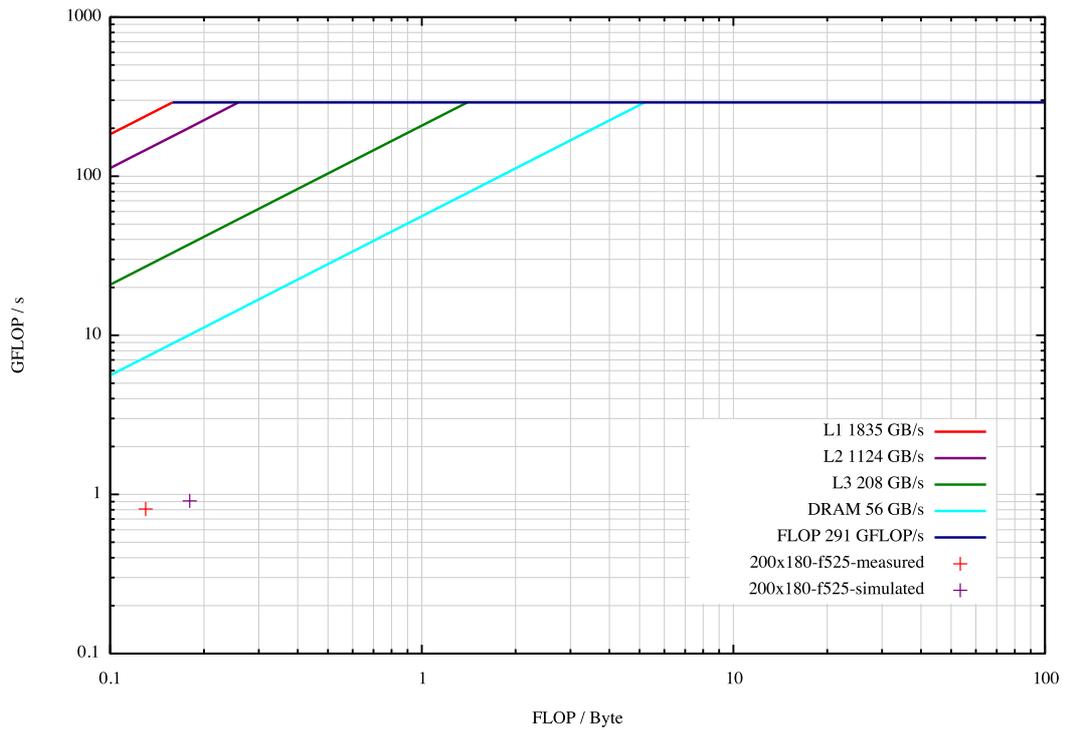

*Figure 25 BiFFT cache-aware roofline on multi-core*

To validate the model different quality metrics are computed:
- The maximum difference is 34.89%;
- The minimum difference is 18%;
- The mean difference is 16.89%;
- The standard deviation is 11.94%.

### 6.1.2.2 Energy model

Energy model is provided in a similar way to one described in Section 5. PKG and DRAM energy coefficients are presented in Table 42 andTable 43 respectively.

| CORES | EPKG [J] | EMPKG [J] | U | S |
|---|---|---|---|---|
| IDLE | 5.53 | | | |
| 1 | 23.28 | 19.08 | | |
| 2 | 35.12 | 19.58 | 0.0427544 | 3.27 |
| 4 | 49.71 | 20.21 | | |

*Table 42 PKG energy coefficients for BiFFT dwarf*

| CORES | EDRAM [J] | EMDRAM [J] | X | Y |
|---|---|---|---|---|
| IDLE | 1.31 | | | |
| 1 | 4.73 | 3.15 | | |
| 2 | 5.42 | 3.55 | 0.1814516 | 0.4144184 |
| 4 | 7.21 | 3.6 | | |

*Table 43 DRAM energy coefficients for BiFFT dwarf*

Table 44 presents comparison between measured and simulated energy for BiFFT dwarfs and different testcases. To validate the model, different quality metrics are computed:
- The maximum difference is 6.08%;
- The minimum difference is 0.23%;
- The mean difference is 8.59%;
- The standard deviation is 2.81%.

| Cores | Testcase | | | | | |
|---|---|---|---|---|---|---|
| | 200x180 fields=20 iter=100 | | 200x180 fields=525 iter=100 | | 800x720 fields=20 iter=100 | |
| | E measured [J] | E simulated [J] | E measured [J] | E simulated [J] | E measured [J] | E simulated [J] |
| 1 | 298.08 | 305.5483945 | 1940.28 | 1944.655154 | 6740.8 | 6784.060006 |
| 2 | 291.27 | 295.5290953 | 1937.2 | 1867.5104 | 6367.35 | 6258.223867 |
| 4 | 300.5533 | 294.2090385 | 1921.705 | 1848.455974 | 6871.2 | 6453.66827 |

*Table 44 BiFFT energy for different testcases - measured vs. simulated*

Figure 26 presents comparison between measured and simulated pairs of time-to-solution and energy-to-solution metrics on a single node, equipped with Intel Xeon E5-2697v3 processor. The current BiFFT implementation suffers from poor OpenMP optimisation, resulting in longer execution time and energy consumption when adding more cores. To save the computational resources, and energy, it makes no sense to use as many cores as delivered within node. In order to decrease time-to-solution, as well as energy-to-solution, one has to run the dwarf in multinode environment, as described in next section.

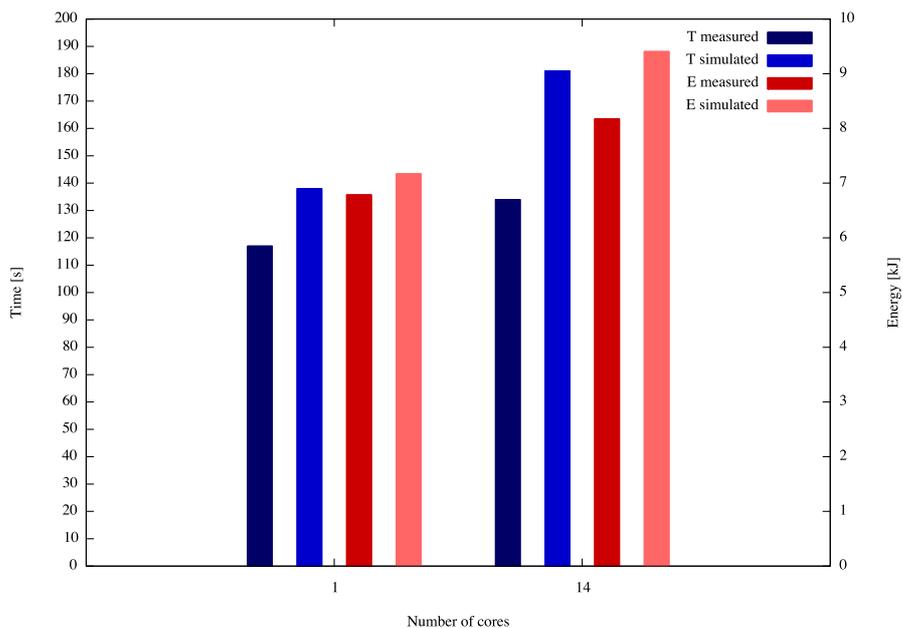

*Figure 26 BiFFT energy comparison on a single node*

#### 6.1.2.1 Multinode

The BiFFT dwarf uses Message Passing Interface to exchange data when running in multinode environment. Example gridpoint distribution is presented in Figure 20. We assume that computational domain is equally distributed over nodes, so that *W* (amount of computations to be performed) and *Q* (amount of data to be sent within all levels of memory within a node) is also equally divided between nodes. We may then use multinode model proposed in Section 5.6.

We use DCworms simulator to project performance and energy efficiency of the BiFFT running in multinode environment, equipped with Intel Xeon E5-2697v3. The projection is made for various nodes-to-cores configurations for the 200x180 testcase: 100 iterations, 525 fields. Figure 27 presents comparison of total execution time between real run (with internode communication) and simulation.

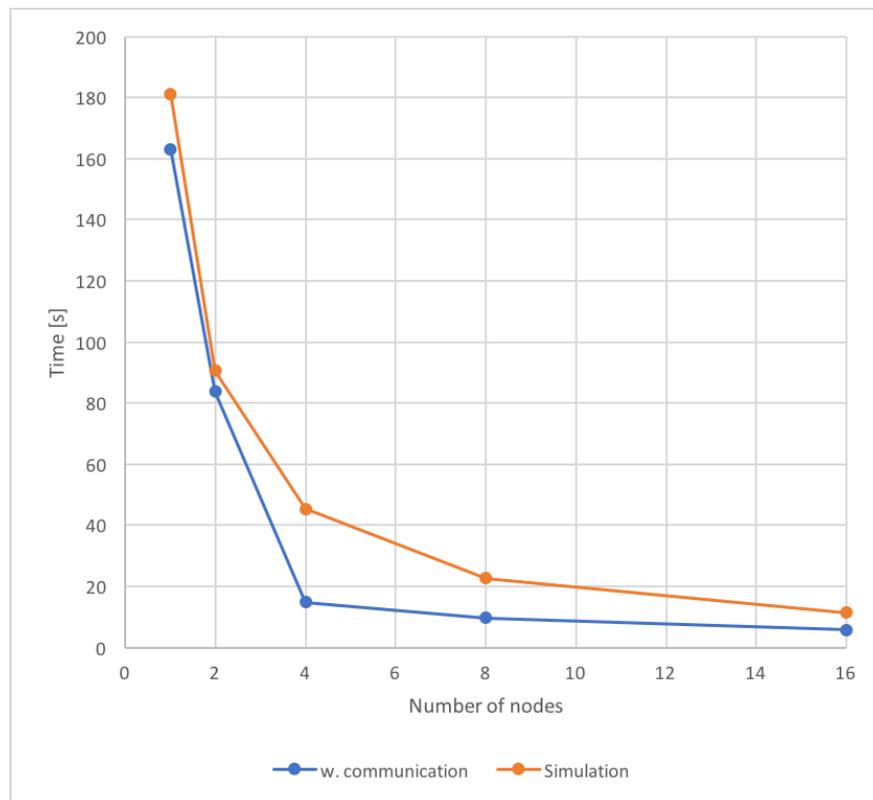

*Figure 27 BiFFT model vs. real run*

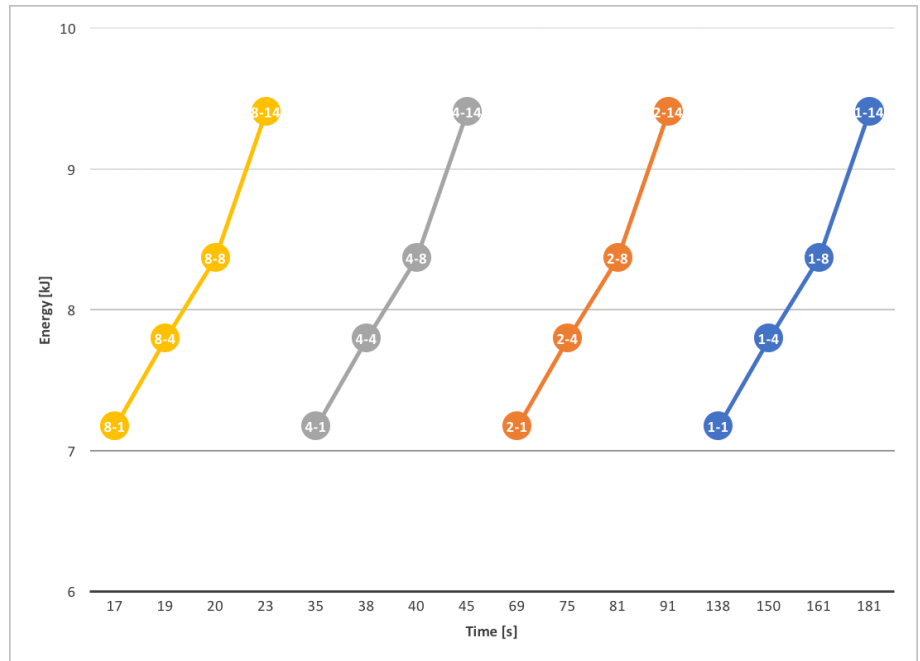

*Figure 28 Time-to-solution and energy-to-solution projection for BiFFT*

Figure 28 provides time-to-solution and energy-to-solution projection for BiFFT dwarf. Same-coloured data points use the same number of nodes. Indices denote the number of nodes (MPI tasks) – left number, and number of OpenMP threads (cores) per task – right number. Increasing number of nodes results in lowering time-to-solution. The amount of work and data to be read/write from/to memory is equally divided between nodes thus time-to-solution decreases equally. However, the energy usage characteristic per node and core remains the same regardless of number of nodes used. That is why dwarf energy consumption remains at more or less the same level. Increasing number of cores results in higher demand for energy, however because of poor OpenMP utilisation the execution time also increases. This projection allows us to select best time-to-solution and energy-to-solution configuration, which is 8 nodes and 1 core. Dwarfs that are suited many-core nodes better will benefit from both, higher number of nodes and cores. For the moment being, DCworms simulator does not take into account balance point, where adding more nodes result in longer time-to-solution and energy-to-solution metrics values. Such situation occurs when computation domain, divided between nodes, becomes too small and communication (including waiting for data to come) takes longer that computations.

### 6.2 Performance and energy efficiency projections of NWP applications

We have selected the ALARO Limited Area Model as a NWP application example. ALARO uses mainly two dwarfs: BiFFT and ACRANEB2. In order to project performance and energy efficiency performance, we introduced possibility to model workflows in DCworms simulator. The weather forecast is run on 200x180 grid and 80 vertical levels. At each time step of ALARO one call to BiFFT and one call to ACRANEB2 occurs. The BiFFT grid is 200x180 with number of fields equal to 640, while ACRANEB2 inputs are KLO=180, KLA=200, KLEV=80. The projection is made on a hybrid architecture. BiFFT is run on many CPU nodes, while ACRANEB2 uses GPU accelerators. We made following architectures available for simulation:
- CPU: Intel Xeon E5-2697v3, Intel Skylake 6148;
- GPU: Nvidia Tesla K20, Nvidia Tesla 2070Q, Nvidia GeForce 970.

We already know that Intel Skylake surpasses Xeon in both, performance and energy consumption. For the GPU and ACRANEB2 dwarf, the time-to-solution does not differ much between different GPU architectures and the best energy-to-solution is delivered with Nvidia Tesla K20. The situation differs when dwarfs are running together. Taking into account energy consumption at idle state, and that Nvidia Tesla K20 uses twice as much power at idle state, time-to-solution is still better for configuration with Nvidia GeForce 970, however energy-to-solution is significantly better with Nvidia GeForce 970, in particular when running on less nodes (see Figure 29).

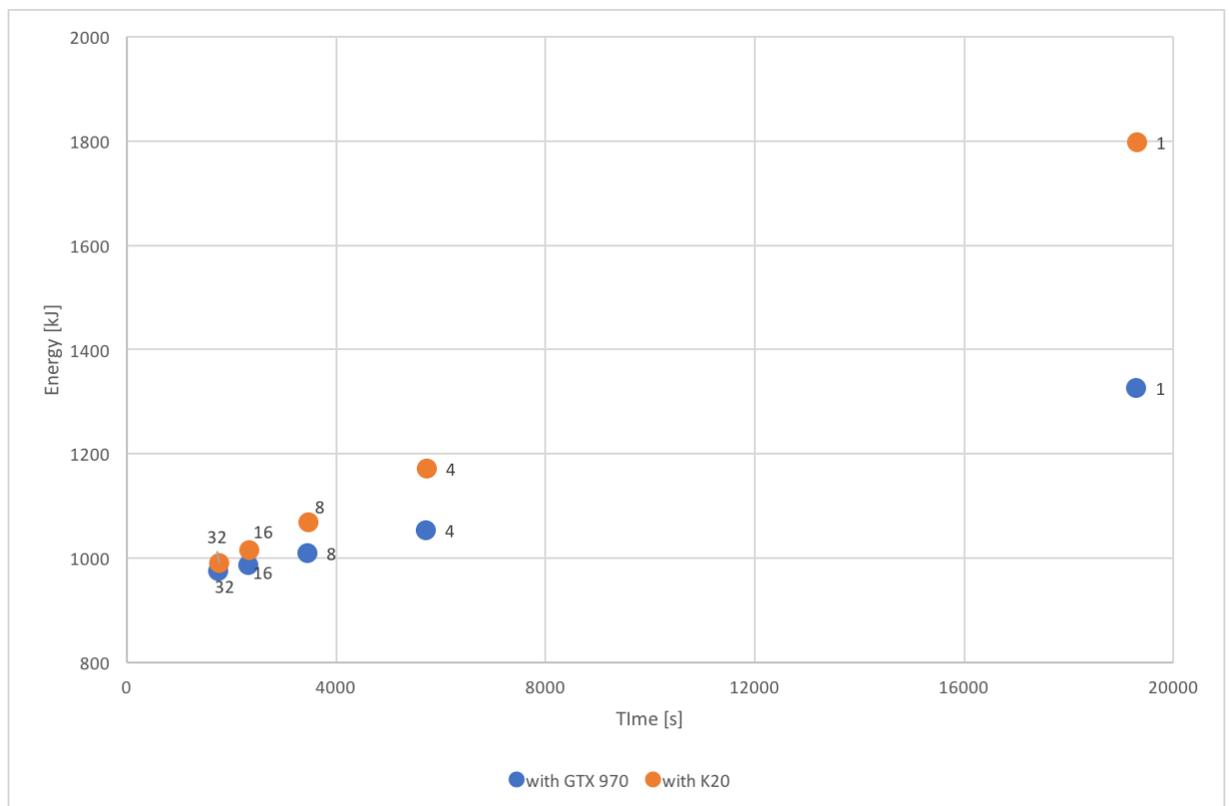

*Figure 29 ALARO time-to-solution and energy-to-solution projection on hybrid architecture (Intel Xeon E5-2697v3 + GPU Nvidia GeForce 970)*

*Indices denote number of nodes.*

DCworms selected the Intel Skylake 6148 and Nvidia GeForce 970 as a best combination that minimises two criteria: time-to-solution and energy-to-solution. Figure 30 presents simulation results of different CPU+GPU configurations on one graph.

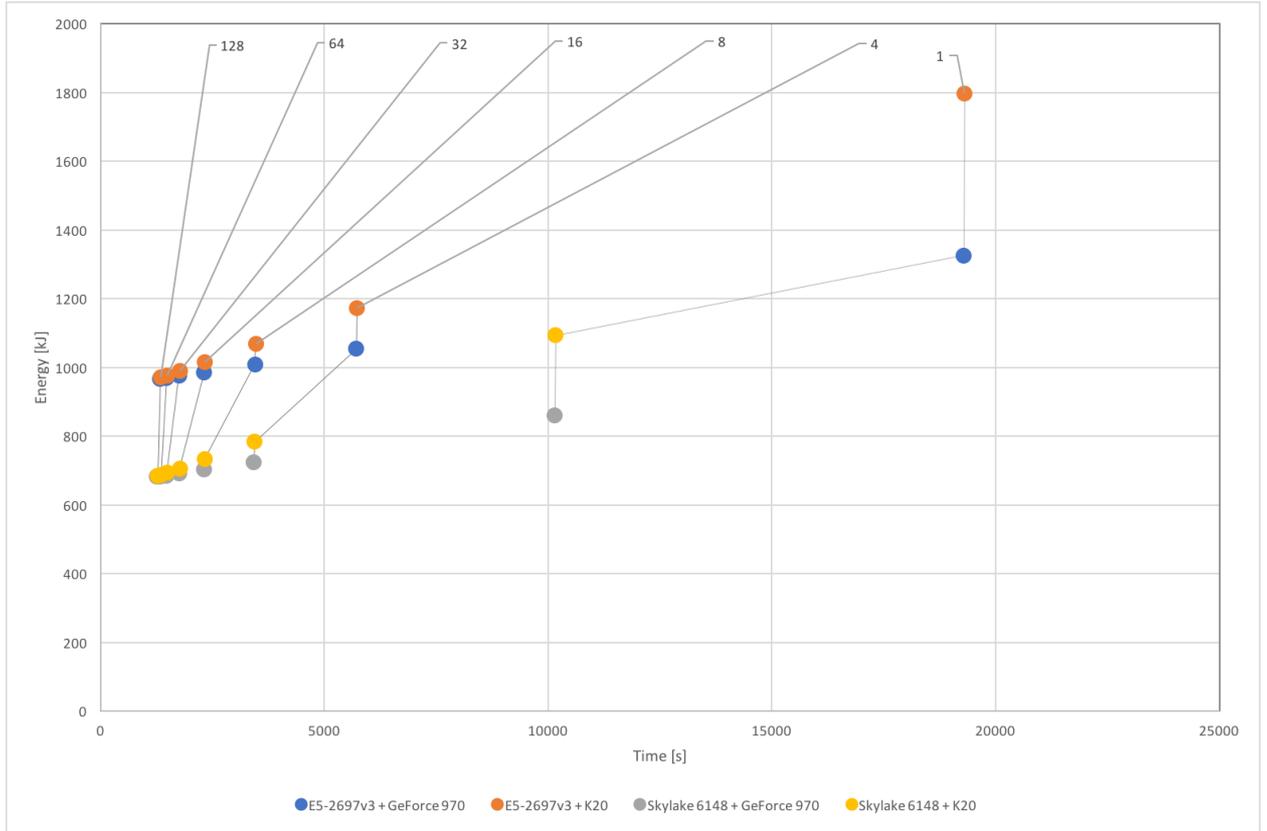

*Figure 30 ALARO time-to-solution and energy-to-solution projection on different architectures*

*Indices denote number of nodes.*

The ALARO LAM operates currently at 2.5km resolution. As described in Deliverable D4.5, it is run on 540x450 grid and 65 vertical levels. To this end, BiFFT should operate at 540x450 grid, while ACRANEB2 on 540x450x65 grid. Unfortunately, this is not possible with tested GPU accelerators, since a single GPU is not capable of handling such grid size with current implementation of ACRANEB2. There should be a multi-GPU version of ACRANEB2 provided to benefit from accelerators at current grid resolution requirements. Please also note, that current ACRANEB2 implementation for GPU accelerators supports only *transt3* subroutine.

# 7  Conclusions & future work

This deliverable presents advances in modelling dwarfs and NWP applications at system scale, taking into account different computation architectures (CPUs and GPUs). The DCworms is capable of modelling cache-aware rooflines, time-to-solution and energy-to-solution metrics, from which a balance point between time and energy may be depicted. Proposed models take into account different memory levels and computation unit performance, as well as energy consumption characteristic. It allows for comparison between different dwarfs and applications runs, where following input variables may be changed and compared: input grid size, model workflow, computation architecture. It also allows to find best hardware configuration for NWP applications.

The DCworms shows its potential to model whole application at system scale, using heterogeneous architectures yet to come. However, there is still plenty of room for improvements. The performance model needs to handle the communication between nodes or rather between MPI processes. As already mentioned in other deliverables, internode communication may have huge impact on overall dwarf performance. This happens especially when there is no possibility to perfectly hide communication by computation. Additional study should concern how different type of internode communication, e.g. point-to-point, all-to-all, affects the performance and how it can be modelled. Deliverable D3.4 presents some information how overall performance may change due to using different communication types.
Another effort should be put on modelling the saturation point of computations, i.e. when adding more cores and/or nodes does not improve scalability, nor time-to-solution.
Another enhancement would be to propose a more detailed model of intra-application (i.e. memory access pattern) and intra-node communication (i.e. OpenMP tasks) to capture scenarios where communication is not perfectly overlapped by computation.
For applications that uses multi-accelerators, e.g. multi-GPU, the multinode model has to be extended.

Last but not least, future work should focus on enhancing energy model. Currently a simplified model was provided, based on available performance and energy counters. As described earlier in this deliverable, state-of-the-art approach is to measure voltages of different hardware components and propose model based on such measurements. Moreover, the currently proposed GPU energy model should be extended with the CPU energy usage as well.

## Document History

| Version | Author(s) | Date | Changes |
|---------|-----------|------|---------|
| *0.1* | *PSNC* | *08/05/2018* | *Draft of the deliverable* |
| *0.2* | *PSNC* | *14/05/2018* | *Energy model* |
| *0.3* | *PSNC* | *15/05/2018* | *Multinode model* |
| *0.4* | *PSNC* | *17/05/2018* | *Further input* |
| *0.5* | *PSNC* | *21/05/2018* | *BiFFT input* |
| *0.7* | *PSNC* | *25/05/2018* | *Complete document draft* |
| *0.8* | *PSNC* | *25/05/2018* | *Addressed reviewers' comments* |
| *0.9* | *PSNC* | *29/05/2018* | *Addressed reviewers' comments* |
| *1.0* | *PSNC* | *29/05/2018* | *Final version* |
| *1.1* | *PSNC* | *10/12/2018* | *Update to address external reviewers' comments on section 4* |

## Internal Review History

| Internal Reviewers | Date | Comments |
|--------------------|------|----------|
| *Michael Baldauf* | *25/05/2018* | *Approved with comments* |
| *Nils Wedi* | *28/05/2018* | *Approved with comments* |

## Effort Contributions per Partner

| Partner | Efforts |
|---------|---------|
| **PSNC** | *13.87* |
| **ECMWF** | *2* |
| *Total* | *15.87* |

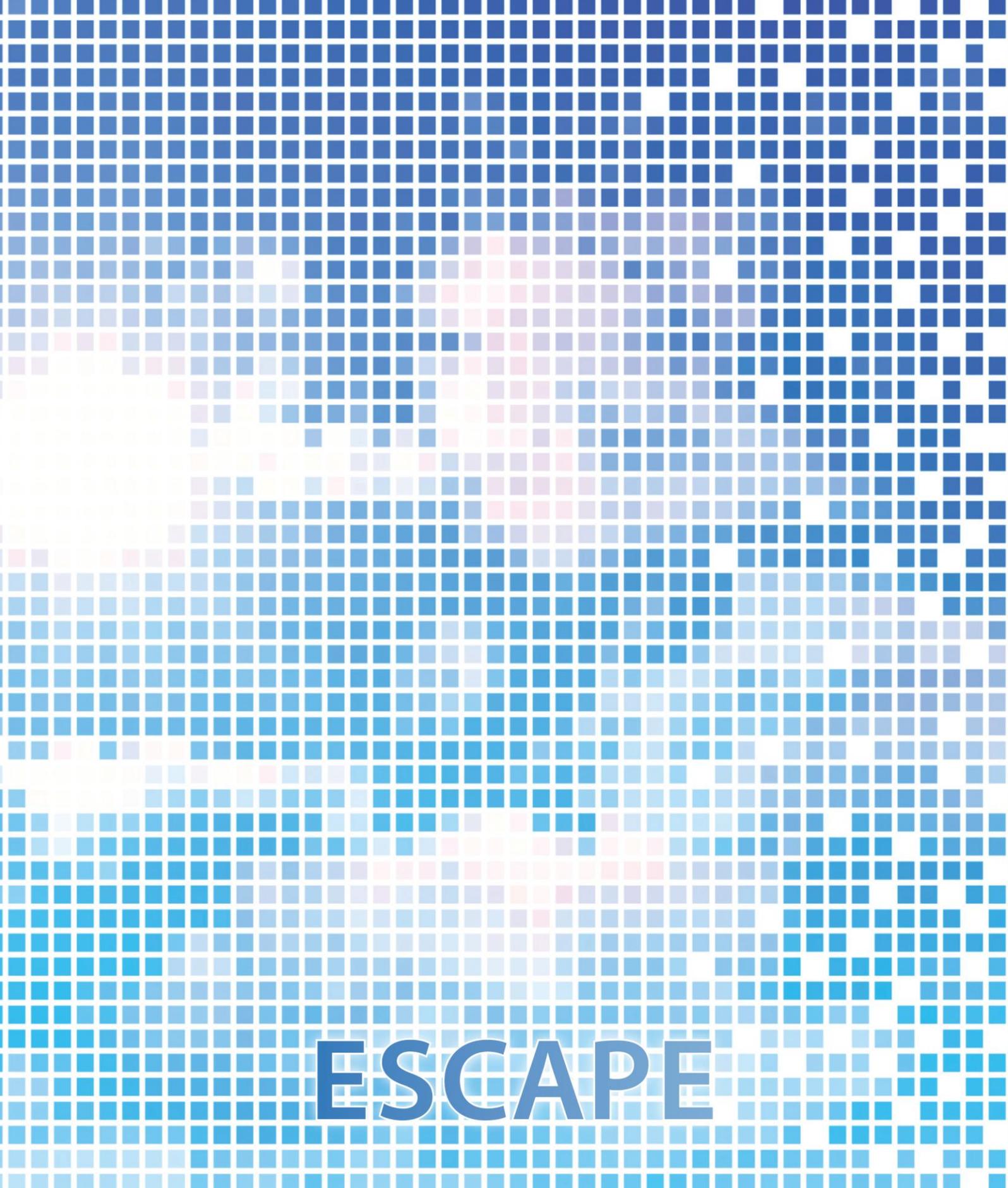